\documentclass{mn2e}

\usepackage[dvips]{graphicx}

\newcommand{\kms}{\mbox{ km s$^{-1}$}}

\begin{document}

\title[Clues for the origin of the metallicity relations]{Clues for the origin of the fundamental metallicity relations. I: The hierarchical building up of the structure }

\author[De Rossi et al. ]{Maria E. De Rossi$^{1,2}$, Patricia B. Tissera$^{1,2}$, Cecilia Scannapieco$^{1,2}$ \\
$^1$ Consejo Nacional de Investigaciones Cient\'{\i}ficas, 
y T\'ecnicas, Rivadavia 1917, (1428) Buenos Aires, Argentina.\\
$^2$ Instituto de Astronom\'{\i}a.
y F\'{\i}sica del Espacio, Argentina, CC67 Suc28, Buenos Aires (1428), Argentina.\\
}

\maketitle
 
\begin{abstract}
We analyse the evolutionary history of galaxies formed in a hierarchical scenario consistent 
with the concordance $\Lambda$-CDM 
model focusing on the study of the relation  between their chemical and dynamical  properties.
Our simulations consistently describe the formation of the structure and its chemical enrichment
within a cosmological context.
Our results indicate that the luminosity-metallicity (LZR) and the stellar mass-metallicity (MZR) 
relations are naturally generated in 
a hierarchical scenario. Both relations are found to evolve with redshift. In the case of  the MZR, 
the estimated evolution is weaker than that deduced from observational works by approximately 0.10 dex.
We also determine a characteristic stellar mass, $M_c \approx 3 \times 10^{10} 
M_{\odot}$, which segregates the simulated galaxy population into two distinctive groups  and which remains unchanged
since $z\sim 3$, with a very weak evolution of its metallicity content.
The value and role played by $M_c$ is consistent with the characteristic mass estimated from the SDSS galaxy survey
by Kauffmann et al. (2004).
Our findings suggest that systems with stellar masses smaller than $M_c$ are  responsible for the evolution 
of this relation at least from $ z\approx 3$. Larger systems are stellar dominated and  have formed more than 50 per cent
of their stars at $z \ge 2$, showing very weak evolution since this epoch.
We also found bimodal metallicity and age distributions from $z\sim3$, which reflects the existence
of two different galaxy populations. 
Although SN feedback may affect
the properties of galaxies and help to shape the MZR, 
 it is unlikely that it will significantly modify   $M_c$ since, from $z=3$ this
stellar mass is found in systems with circular velocities larger than $100 \kms$. 

\end{abstract}


\section{Introduction}
\label{intro}

Chemical abundances  could provide important clues on  the evolutionary history 
of galaxies (e.g., Freemann \& Bland-Hawthorn 2002) since  they are the  result of the joint action of several physical processes 
such as Supernova (SN) feedback (Larson 1974; Larson \& Dinerstein 1975; White \& Rees 1978; 
Kauffmann \& Charlot 1998; Somerville \& Primack 1999; Cole et al. 2000; Springel \& Hernquist 2003), gas inflow  (Pagel 1997),
 mergers and interactions,  among others. 
And, although it is not easy to disentangle the effect of each individual process, it may be possible to constrain 
galaxy formation models by studying how chemical and dynamical properties of galaxies evolve as a function of cosmic epoch (Pettini 2006) and confronting them with observations.

There are strong evidences for the existence of a good correlation between  the luminosities of local galaxies and 
their chemical abundances  which show that metallicities tend to increase with galaxy luminosity. 
In addition, recent studies have suggested that the LZR evolves with redshift so
 that, at a given metallicity, galaxies were brighter  in the past (Pettini et al. 2001; Kobulnicky et al. 2003; Kobulnicky \& Kewley 2004; Lamareille et al. 2004; Maier et al. 2004; Shapley et al. 2004).
An evolution in both the zero point and the slope is suggested, indicating a differential evolution in the
chemical content of the systems as a function of luminosity.  In particular, 
in agreement with previous works, Kobulnicky \& Kewley (2004), 
found that the largest change from $z \sim 3$ is driven by  faint galaxies, while from $z \sim 1$, they detected an 
increase in the metallicity level of $\sim 0.14$ dex. Interestingly, Erb et al. (2006a) showed that 
the large uncertainties present
in the determination of the LZR at high redshift,
introduced mainly by the variations in the mass-to-light ratios, make it very difficult
to draw a robust conclusion on its evolution.
 
Luminosities are usually used as a surrogate of stellar mass because of the difficulty in measuring the latter.
However, it is now accepted that the correlation between metallicity and stellar mass is more fundamental.
This  correlation  was first reported by Lequeux et al. (1979) in a study of irregular galaxies.
Recently, Tremonti et al. (2004, hereafter T04) confirmed this correlation on a statistical basis by 
determining a tight correlation between gas-phase metallicity and stellar mass, extended
 over a factor of ten in metallicity and three decades in stellar mass,
 for a sample of $\approx 30000$ local star-forming galaxies in  the Sloan Digital Sky Survey (SDSS).
 The stellar MZR shows a linear growth between  $10^{8.3} {\rm M_{\odot}} h^{-1}
$ and $10^{10.35} {\rm M_{\odot}} h^{-1}$, flattening for larger stellar mass. T04 interpreted this behaviour as the efficient action of galactic winds over systems with masses lower than $\sim 10^{10.35} {\rm M_{\odot}}h^{-1} $.
 This stellar mass  has been previously identified   as a characteristic mass for galaxy 
evolution (Kauffmann et al. 2004). Furthermore,  Gallazzi et al. (2005, hereafter G05) analysed stellar masses, light-weighted ages and stellar metallicities for a sample of galaxies of the SDSS finding similar trends  to those obtained by T04. G05 observed that both age and metallicity tend to increase with stellar mass, showing a rapid growth at intermediate masses and a gradual
 flattening  above  $M_*\sim 10^{10.35} {\rm M_{\odot}} h^{-1}$.
Recently, Savaglio et al. (2005) estimated a relative evolution of 0.10-0.15 dex at $z\approx 0.7$ with
respect to the local MZR of T04 while  Erb et al. (2006a) found a relative evolution
of $\approx 0.30$ dex for  the MZR of galaxies at $z\approx 2.5$.

The understanding of the origin of the LZR and MZR within the context of the current cosmological
paradigm can provide more stringent constrains for galaxy formation models since these correlations are produced
by the intervening action of different physical processes.
 In this scenario, galaxies formed by the hierarchical aggregation of
substructures which can affect the internal distribution of angular momentum and mass in galaxies, modifying
the star formation rate and the metal production and distribution. SN energy feedback is also expected to
play a main role by contributing to establish a self-regulated star formation activity and to trigger powerful
outflows. Finally, the interaction between a galaxy and its environment (i.e. groups or
clusters) may strip gas, starving galaxies and quenching the star formation activity and the synthetization of
new chemical elements.
 
Numerical simulations which can describe the non-linear evolution of the matter and
its chemical enrichment self-consistently have proved to be an adequate tool to tackle galaxy formation 
(Mosconi et al. 2001; Lia et al. 2001; Kawata \& Gibson 2003).
 In particular,  Tissera, De Rossi \& Scannapieco (2005) used chemical hydrodynamical simulations to 
 show that galaxy-like systems in 
the concordance $\Lambda$-CDM model followed the general trend of the observed  MZR, 
predicting an evolution in zero point and slope.
 The authors also claimed the need for strong SN galatic winds to decrease the fraction of
metals locked into stars, specially in low mass systems. Although SN energy feedback has been 
regarded as a crucial physical mechanism to regulate the star formation activity in galaxies and to
produce metal-loaded, gas outflows capable of enriching the intergalactic medium, its modelling within
hydrodynamical simulations is still controversial.  For this reason,
in this work we analyse in more detail  the origin of both the LZR and the MZR and study the properties of the simulated
 galaxies with the aim at improving our understanding of the evolution of both relations in a hierarchical scenario,
focusing  on the role of dynamics.  A secondary goal of our work is to clarify where and how SN energy feedback
is required. A full study including SN energy feedback will be carried out in the future when the SN model
of Scannapieco et al. (2007, in preparation) gets fully tuned to be used in cosmological simulations.

This paper is organized as follows. Section 2 summarizes the main characteristics of the simulations. 
Section 3 provides a description of the simulated galaxies and an analysis of  their mean astrophysical properties.
 Section 4 and Section 5  discuss
the simulated fundamental metallicity relations and their evolution. Section 6 investigates the origin of the characteristic
stellar mass. And Section 7 summarizes our main results.

\section{Numerical simulations}

We  run numerical hydrodynamical simulations consistent with  the concordance $\Lambda$-CDM universe with
$\Omega =0.3, \Lambda =0.7, \Omega_{b} =0.04$  and $H_{0} =100 h^{-1}$ km s$^{-1} {\rm Mpc}^{-1}$ with  $h=0.7$.
  These simulations were performed by using the chemical   {\small GADGET-2}, which includes a
treatment for metal-dependent radiative cooling, stochastic star formation  and chemical enrichment (Scannapieco et al. 2005).
 
The chemical   {\small GADGET-2} code describes the  enrichment by Type II and Type Ia  Supernovae (SNII and SNIa)
 according to the chemical yield prescriptions of Woosley \& Weaver (1995) and Thielemann, Nomoto \& Hashimoto (1993), respectively. Instantaneous thermalization of the supernova (SN) energy has been assumed in this work. 
Previous works have shown that the injection of SN energy directly into the internal energy of the gas
has no impact on the dynamics, so in our model SN energy feedback is inefficient
 (see Marri \& White 2003 for a comprehensive review).
We adopted a standard Salpeter Initial Mass Function with a lower and upper mass cut-offs of 0.1 $\rm M_{\odot}$ and 40 $\rm M_{\odot}$, respectively.  For SNIa we assumed a time-delay for the ejection of material randomly chosen within 
 $[0.1,1]$ Gyr. We assumed instantaneous recycling condition for SNII.
 Metals are distributed within the neighbouring gas particles weighted by the 
smoothing kernel. For more details on the chemical model, the reader is referred to Scannapieco et al. (2005).

The simulated volume corresponds to a cubic box of a comoving  10 Mpc $h^{-1}$ side length. We 
 ran two experiments resolved initially with $2\times 160^3$ (S160) and $2 \times 80^3$ (S80)  particles, respectively.
 Hence, the mass resolutions considered are of  $2.17 \times 10^7 {\rm M_{\odot}} h^{-1}$ (S80) and $2.71 \times 10^6 {\rm M_{\odot}} h^{-1}$ (S160) for the gas component, and  $1.41 \times 10^8 {\rm M_{\odot}} h^{-1}$ (S80) and $1.76 \times 10^7 {\rm M_{\odot}} h^{-1}$ (S160) for dark matter.

\section{Simulated Galaxies}

The identification of the simulated galaxies  was carried out by combining the   friends-of-friends technique  
and a contrast density criterion which selects virialized structures as
described  by  White, Efstathiou \& Frenk (1993).
 In order to diminish numerical resolution problems, we only analysed those systems with 
virial masses greater than  $\sim 10^9  {\rm M_{\odot}} h^{-1} $. 
In S160, the analysed simulated galaxies are then  required to have more than 2000  particles within
the virial radius, while systems in S80 have at least 200 particles.
 Hence, our results are  valid for galaxies with masses larger than this cut-off mass. The comparison between results from S80 and S160 will
allow us to assess the effects of numerical resolution on the results.

In order to perform a better comparison with observations, 
the star formation histories as well as mean dynamical and chemical properties of the simulated galaxies are
 estimated at the optical radius $r_{\rm opt}$, defined as the one which encloses 83 per cent of
 the baryonic mass of the system. 
The typical   $r_{\rm opt}$ for the simulated  systems identified at $z=0$ ranges  within $[2,20]$ kpc $h^{-1}$,
 with the mean value at $\sim 6$ kpc $h^{-1}$.
For the sake of comparison with observations, we also estimate the aperture radius $r_{\rm ap}$ as the one
that encloses 33 per cent of the baryonic mass of the system. This radius will be used to estimate
 mean abundances for a direct comparison with  observational data.

\subsection{Chemical properties}

Our  simulations  provide the  abundances of several individual chemical elements  such as
$^{16}$O, $^{56}$Fe, H, etc.,  present in baryons as a result of stellar evolution.
 In particular,  we can  directly examine indicators such as the mean oxygen abundance, 12+log(O/H). However, the confrontation
of the simulated results with observations may be tricky since, besides several  observational difficulties
related to the determination of  chemical abundances, some observed estimators could be biased
 toward high star formation regions
(e.g. estimations for HII regions) or the central part of galaxies (e.g. estimations from SDSS).
Hence, depending on the observational data chosen to carry out a comparison,
 aperture effects must be taken into account, for example.

We define the  mean mass-weighted O/H indicator  as

\begin{center}
\begin{equation}\label{eq:W_OH}
<12+{\rm log(O/H)}>_{\rm W} = 12+ {\rm log} 
\Bigg(\frac{\sum_{i=1}^{ N_{\rm p}} {\rm O_i} }
{ \sum_{i=1}^{ N_{\rm p}} {\rm H_i}}\Bigg),
\end{equation}
\end{center}

\noindent where ${\rm O}_i$ and ${\rm H}_i$ are the oxygen and hydrogen atomic abundances  in  particle $i$ and
$N_{\rm p}$ corresponds to the total number of particles within an optical radius (or the aperture radius).
This indicator can be easily estimated for the gas and stellar components in a  given simulated galaxy.

\begin{figure*}
\begin{center}
\vspace*{0.5cm}\resizebox{8.5cm}{!}{\includegraphics{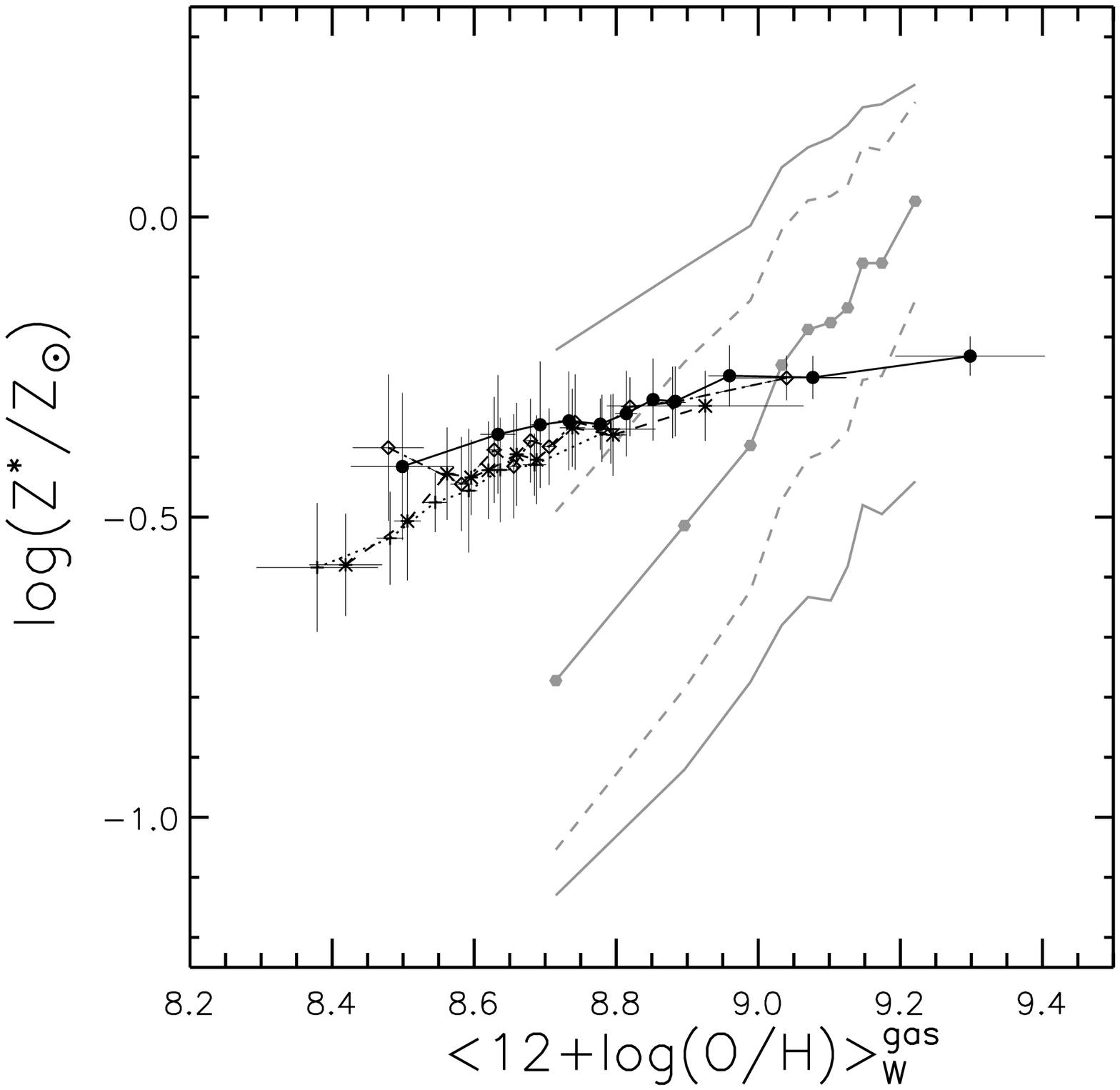}}\hspace*{-0.2cm}%
\vspace*{0.5cm}\resizebox{8.5cm}{!}{\includegraphics{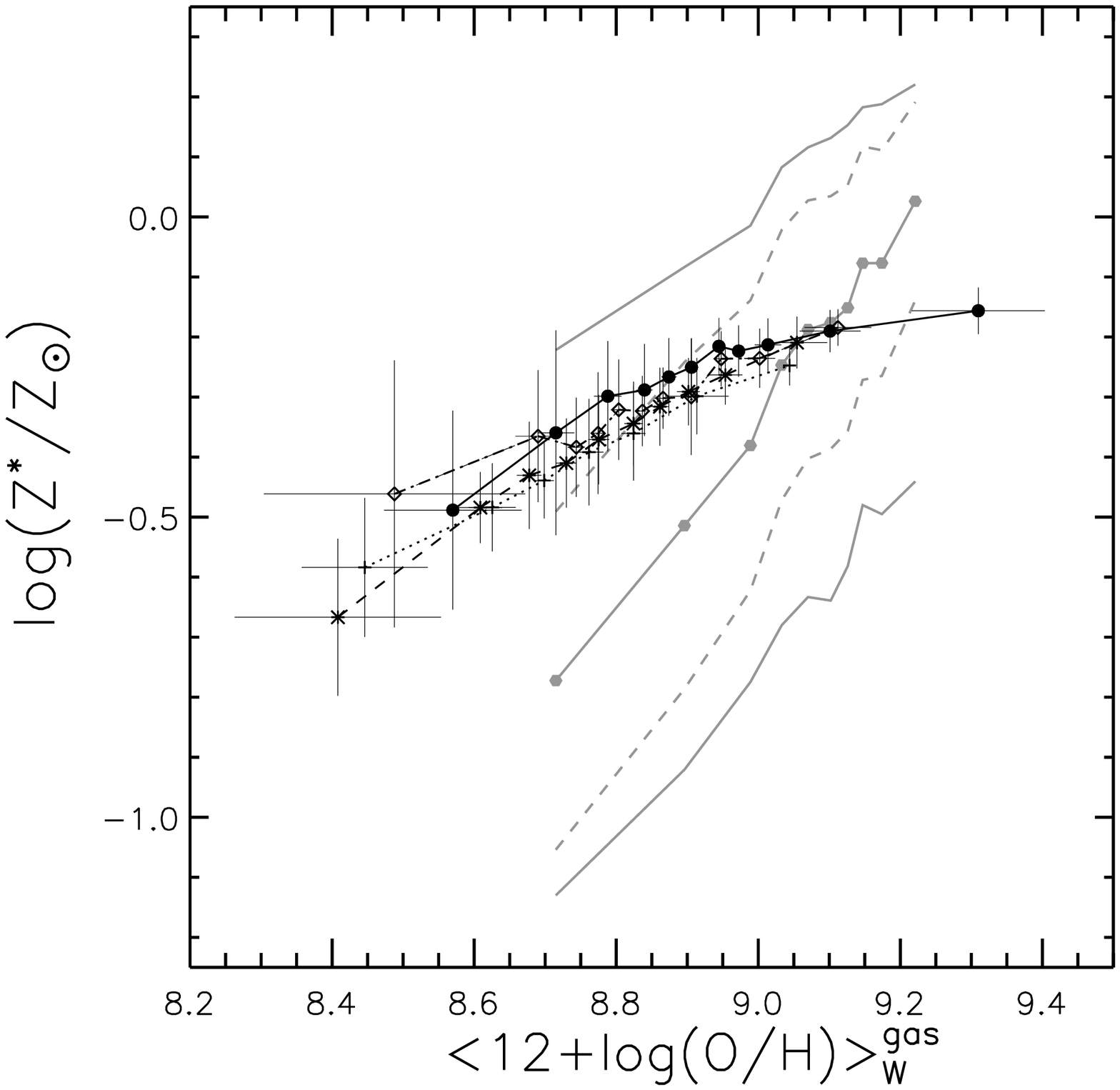}}\hspace*{-0.2cm}%
\end{center}
\caption[12+log(O/H) en funci\'on de $M_{\rm u}$, $M_{\rm g}$ y $M_{\rm r}$ a $z=0$]
{Mean stellar metallicity $Z^*$ and
gas O/H abundance for simulated
galaxies at $z=3$ (dotted line), $z=2$ (dashed line), $z=1$ 
(dotted-dashed line) and $z=0$ (solid line), considering the mass enclosed by 
${r_{\rm opt}}$ (left panel) and
$r_{\rm ap}$ (right panel). The error bars denote
the rms deviations about mean values.
The grey filled circles show the median stellar metallicity for the 
sample of SDSS galaxies analysed by Gallazzi et al. (2005), while the outer
solid lines show the corresponding 16th and 84th percentiles. The dashed lines
indicate the mean 68 per cent confidence range in the stellar metallicity estimated.
}
\label{fig:ZsvsOHg}
\end{figure*}

G05 have recently obtained
a positive correlation between  O/H abundance of the gas phase and 
the stellar metallicity $Z^*$ (i.e. 
total mass of all elements heavier than helium
over the total stellar mass) for a sample of star-forming galaxies
of the SDSS. In Fig. \ref{fig:ZsvsOHg} (left panel), we show this relation
for the simulated galaxies at different redshifts. Although the simulated abundances
are within a range of values consistent with those obtained by G05 at $z=0$,  the simulated
relation is shallower than
that estimated for the SDSS galaxies. 

Taking into account that the SDSS provides information of the central regions of galaxies defined
by the aperture angle, 
in order to evaluate
at which level aperture effects could be responsible for
this discrepancy, we estimated the stellar metallicity $Z^*$ and 
O/H gas phase abundance within
$r_{\rm ap}$. The results are shown in Fig. \ref{fig:ZsvsOHg} 
(right panel).  We see that, limiting our metallicity estimations to
the central region of galactic systems leads to a steeper correlation
in better agreement with observations.  Nevertheless, there is still
a discrepancy which 
may be owing to the lack of strong energy feedback in our
numerical model, which might have a direct impact on the chemical enrichment of 
the matter, principally, for low mass systems (see for example Scannapieco et al. 2006).
 
Regarding the evolution of this relation with redsfhit, from Fig. \ref{fig:ZsvsOHg} 
(right panel), we can see that, at a given gas phase metallicity and albeit the high dispersion,
 galactic systems  tend to have systematically
more enriched star at low redshift.
Our simulations predict a mild evolution (i.e. less than 0.10 dex) in the correlation
between stellar metallicity $Z^*$ and gas phase abundance from $z=3$ to $z=0$ which is, nevertheless, within the estimated standard
dispersion.

Finally, as noted by Tissera et al. (2005),  the abundances of the gas phase have a large dispersion which
introduces important noise to  the estimated relation. For this reason, these authors decided to use
the abundances estimated from the stellar populations to assess the level of evolution in the
 MZR. In this paper, we use both indicators for the comparison with observations. But chemical
indicators estimated by using  the stellar populations will be used for the rest of the discussions since the signals and
noise is significantly smaller.

We tested that numerical resolution is not strongly affecting these determinations by analysing the results from 
 S80 run which are found to be consistent with those of S160. 
Hereafter, for the sake of simplicity we will show only the results for S160 but simulated galaxies in S80 determine
similar relations and results.

\subsection{Gas fractions and masses}

Gas fractions in galaxies are the result of the combined effects of the rate of transformation of gas into stars, gas infall and outflows.
Mergers and interactions may help to increase the rate of accretion and  
to strip material from the systems, playing a role
in the regulation of the star formation activity.
Local observations show an anticorrelation between gas fraction and luminosity 
(e.g. Boselli et al. 2001) or stellar masses (Brinchmann \& Ellis 2000).
 Recently,
Erb et al. (2006b) reported hints for a similar relation at $z \sim 2-2.5$. Moreover,
these authors found a mean $M_{\rm gas}/M_{\rm bar} \sim 0.50$ while T04 measured  a mean
of 0.20 for galaxies at $z\sim 0$. This trend illustrates the transformation of gas into stars with time.
Since in our simulations, the gas reservoirs in the simulated galaxies are determined by the interplay between
gas infall and star formation activity, which are both regulated by the hierarchical assemble of the structure, it
is interesting to analyse at what extent the mentioned observational results can be reproduced
by our simulations.

\begin{figure*}
\begin{center}
\vspace*{0.5cm}\resizebox{8.5cm}{!}{\includegraphics{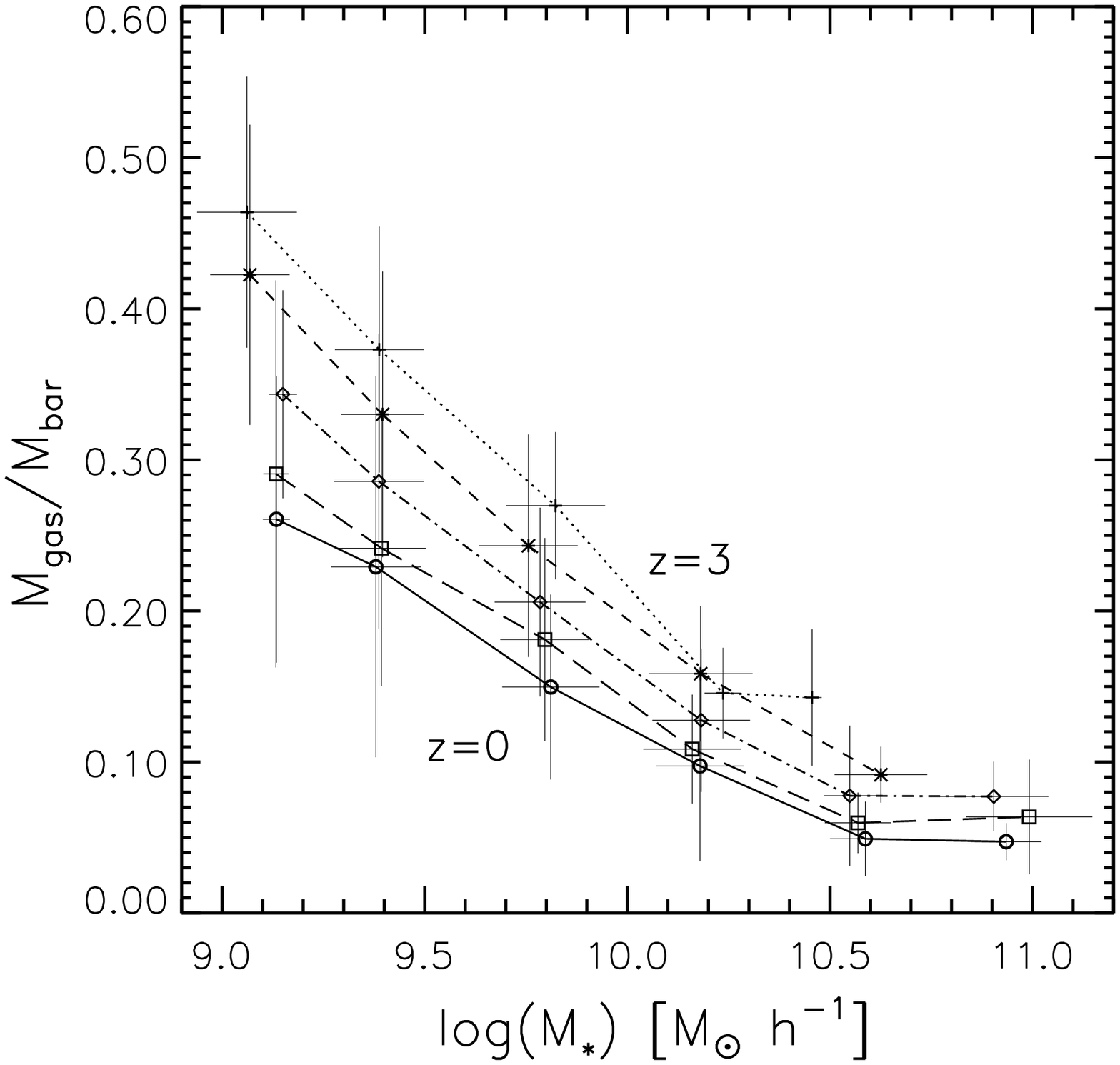}}\hspace*{-0.2cm}
\vspace*{0.5cm}\resizebox{8.5cm}{!}{\includegraphics{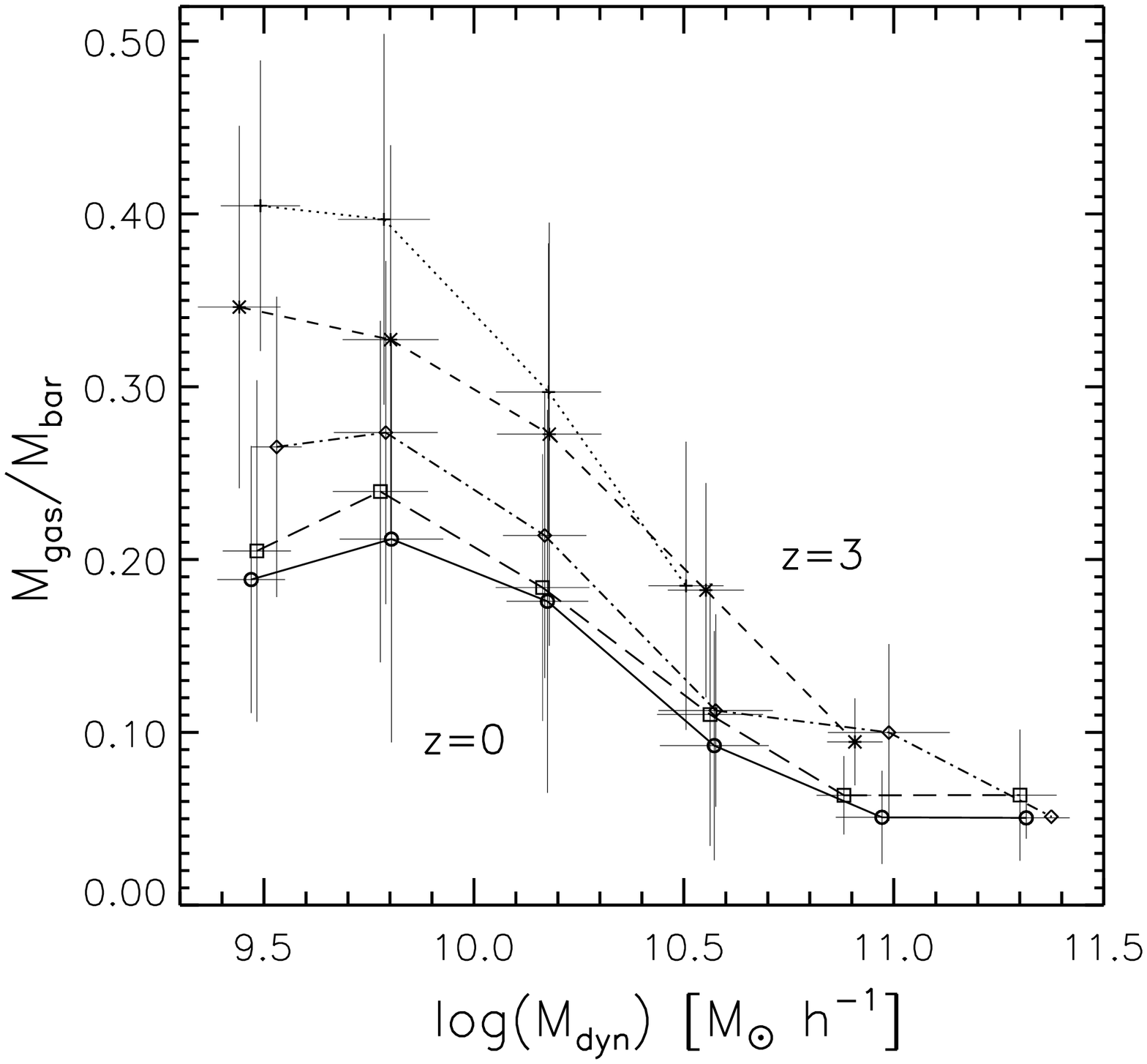}}\hspace*{-0.2cm}
\end{center}
\caption
{Gas mass normalized to total baryonic mass within the optical radius as
a function of stellar mass (left panel) and dynamical mass
(right panel) for simulated galaxies  at
$z=3$ (dotted line), $z=2$ (dashed line),
$z=1$ (dotted-dashed line),
$z=0.5$ (long-dashed line) and
$z=0$ (solid line).
The error bars correspond to the rms deviations about the mean values.}
\label{gasfrac}
\end{figure*}

Hence, we estimated the gas fraction in the simulated galaxies as a function of stellar mass  from $z\sim 3$ to $z\sim 0$
as displayed in 
  Fig.~\ref{gasfrac} (left panel).
The fraction of gas increases for higher redshifts,  with the largest changes for
the smallest systems. At the low stellar mass end, the dispersion in 
the gas fraction is quite large,  indicating
that  these systems have different evolutionary histories.
Galactic systems with $M_{*} > 10^{10} {\rm M_{\odot}} h^{-1}$
have already consumed most of their gas into stars at $z > 3$.
While at $z= 0$, we found   mean $M_{\rm gas}/M_{\rm bar}$ varying between $\sim 0.20$ and $\sim 0.05$,
at $z\sim 2$  mean values are in the range  $0.10-0.40$. These latest gas fractions are lower by  a factor of $\sim 2$
than the mean
one reported by Erb et al. (2006b).
 This finding suggests again the need for strong SN winds to
regulate the transformation of gas into stars as the systems are assembled.
The lack of strong SN winds results in our findings being upper
limits for the fraction of stars that systems can form at different redshifts
 as a result of their dynamical
evolution in hierarchical scenarios.
However, we also note that Erb et al. (2006b) assumed a Chabrier IMF  while in our simulations,
 we adopted a Salpeter IMF, fact that leads to an 
overestimation by  up to a factor of $\sim 2$ of the fraction of baryons locked into stars. 
The different IMFs could account for these differences between the simulated and observed gas fractions. 

 According to Fig.~\ref{gasfrac} (left panel), there is a variation of a factor of 5 in gas fraction
over a change in 2 dex in stellar mass.
 It will be interesting to see if the same trend holds if  dynamical masses are used instead of stellar masses.
This is indeed the case, as shown in  Fig.~\ref{gasfrac} (right panel), although
the relation is flatter for lower dynamical masses owing to the contribution of systems with different
 gas fractions at a given dynamical mass.
Dynamical masses are calculated taking into account all baryons and dark matter within an optical radius.

\begin{figure}
\begin{center}
\vspace*{0.5cm}\resizebox{7.5cm}{!}{\includegraphics{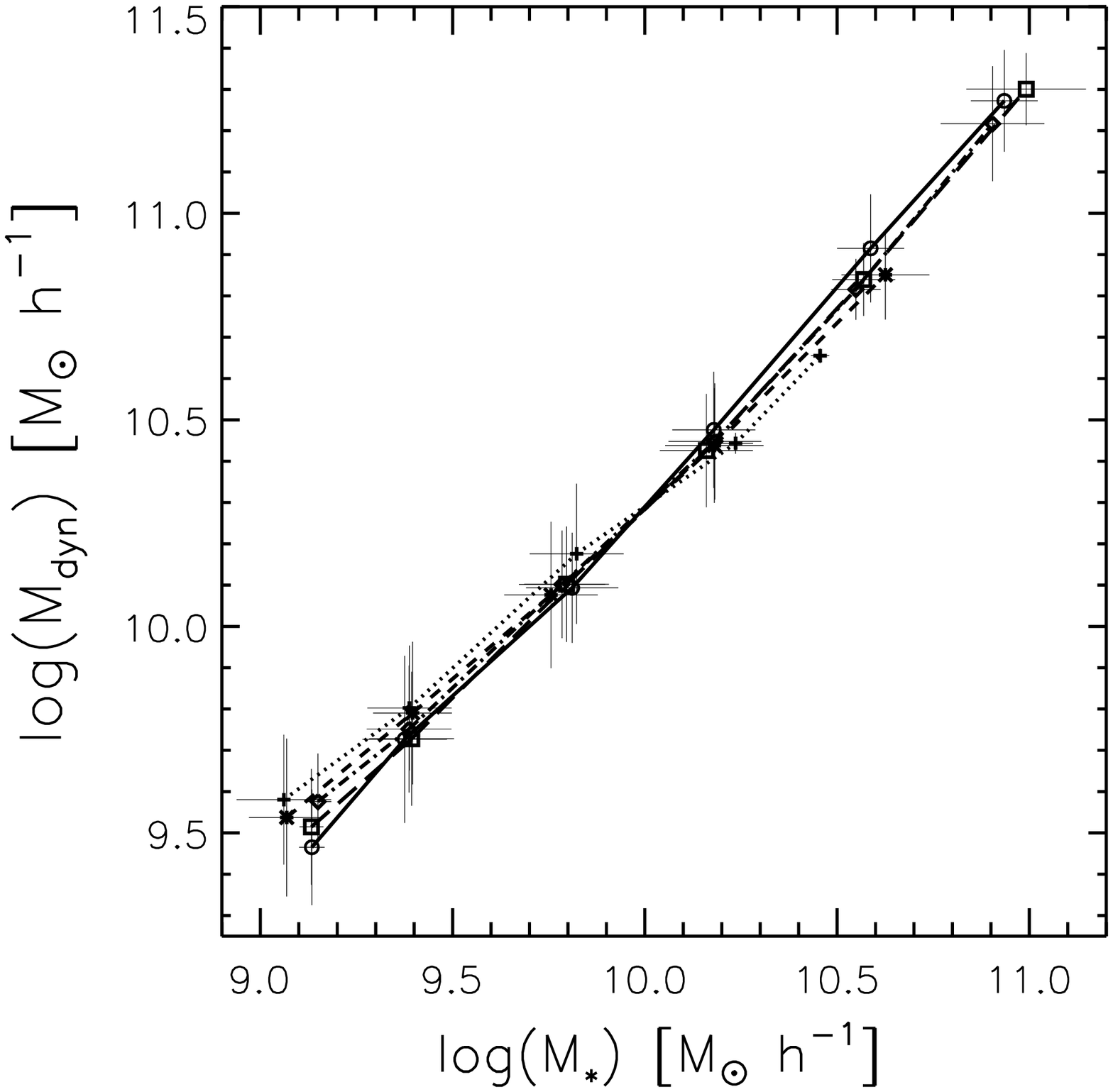}}\hspace*{-0.2cm}\\
\vspace*{0.5cm}\resizebox{7.5cm}{!}{\includegraphics{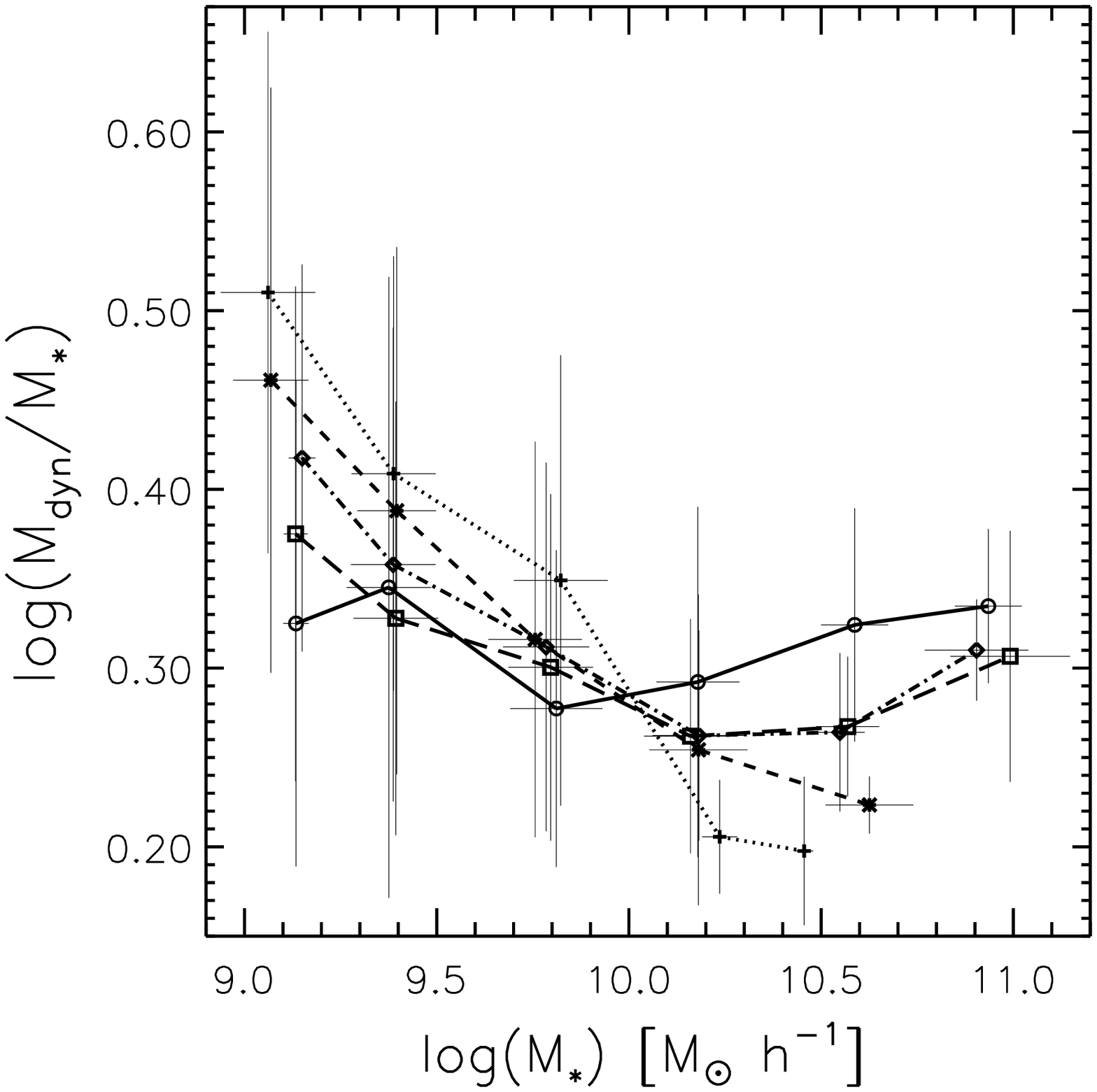}}\hspace*{-0.2cm}
\end{center}
\caption
{Dynamical mass versus stellar mass  (upper     panel) and the ratio between dynamical and stellar mass as function 
of stellar mass (lower panel) estimated at the same redshifts quoted in Fig.~\ref{gasfrac}.
The error bars correspond to the rms deviations about the mean values.}
\label{gasfracmass}
\end{figure}

Observations predict a good correlation between stellar and dynamical masses (e. g. Brinchmann  \& Ellis 2000).
We found this tight correlation  from $z \approx 3$,
as it can be seen from Fig. ~\ref{gasfracmass} (upper panel).
There is a small change in the slope of the relation from $z=3$ to $z=0$.
This behaviour can be also understood from 
 Fig. ~\ref{gasfracmass} (lower panel) where we show the ratio between dynamical and stellar mass as a function of 
stellar mass. 
On average, the hierarchical growth of the structure produces systems with dynamical
masses a factor of 2.0-2.5 higher than their mean stellar mass since $z\sim 3$. This trend is consistent
with that recently reported by Erb et al. (2006b) from observations of galaxies at $z=2.0-2.5$.
In the simulations, the anticorrelation is produced by the fact that smaller systems have a larger fraction of baryons
in form of gas at all redshifts. 
For lower redshifts, the slope of the relation changes because small systems continuously transformed their gas components into stars
while
larger systems increase  slightly their dark matter content within an optical radius.
This last behaviour was determined by estimating the fraction of each mass component (i.e. gas, stars and dark matter)
 within an optical radius for all systems as a function of redshift.

\section{The Luminosity-Metallicity Relation}

The magnitudes for our simulated galaxies have been estimated by applying  the technique described
by Tissera et al. (1997) which combines the information on the stars generated by the simulations and
 population synthesis models.
In particular, we use  the  models of Bruzual \& Charlot
(2006, BC06, in preparation), which enables the determination of the  spectral properties for
stellar populations with different levels of enrichment.
The integrated-light of simulated galaxies were estimated by convolving
the age and chemical 
content of each simulated stellar population (represented by a star particle) with the corresponding synthetic spectrum generated
by  BC06 models, according to 
\begin{equation} \label{eq:flujosimu}
F_{\lambda}=
\sum_{\rm j}  m_{\rm j} \, F(\lambda,Z_{\rm j},t_{\rm j}) \, f_{\rm ext}(\lambda,\tau_{V},t_{\rm j})
\end{equation}
where $m_{\rm j}$, $t_{\rm j}$ and $Z_{\rm j}$ correspond to the mass,
the age and the metallicity of star $\rm j$ respectively.
The function $F$ depicts the flux at a wavelength $\lambda$
for a star with a given age and metallicity, while $f_{\rm ext}$
is introduced in order to describe the attenuation by dust and
is defined following the work of Charlot \& Fall (2000) as
\begin{equation}\label{eqn:extincion}
f_{\rm ext}(\lambda,\tau_{V},t_{\rm j})= e^{- \tau_{\lambda}(\lambda, \tau_{V},
t_{\rm j})}
\end{equation}  
where $\tau_{V}$  is the total optical depth in the V band affecting young
stars.  The relation between  $\tau_{\lambda}$ and $\tau_{V}$ is given
by the extinction curve, which describes the attenuation due to photons
 emitted in all directions by stars of age $t_{\rm j}$ in a galaxy.
Charlot \& Fall (2000) consider a scenario in which stars form in interstellar
clouds and after $10^7$ yr emigrate towards the interstellar medium, 
deriving a extinction curve with the form
\begin{equation}
\tau_{\lambda} (\lambda,\tau_{V},t) =
\left\{ \begin{array}{ll}
\tau_{V} \, {(\lambda/5500 {\mathrm \AA})}^{-0.7}
& \textrm{if $t \le 10^{7}$ years} \\
\mu \, \tau_{V} \, {(\lambda / 5500 {\mathrm \AA})}^{-0.7}
& \textrm{if $t > 10^{7}$ years}\\
\end{array} \right.
\end{equation} 
where $\mu$ is an adjustable parameter which represents the fraction
of the total optical depth contributed by the interstellar medium.

In this work we consider a random distribution of $\mu$
along the range $[0,0.6]$ as suggested by  BC06. 
For $\tau_{V}$, we use recent observational  results which  suggest that it may be a function of the metallicity of
the galaxy. To estimate this last correlation we made a 
polynomial fit to 
the observed correlation obtained from
the SDSS (Brichmann, private communication) and assumed Gaussian
uncertainties.
With this model we determined the spectral properties of galaxies
and derived colours and magnitudes.

At $z=0$ we found that the simulated galaxies determine a LZR relation in general good agreement with observations as it
can be appreciated from Table \ref{tab:LMR_z0} where we show the results of the linear regression fits to the simulated relation
in the B-band. For comparison we also include  the corresponding fit for the SDSS galaxies 
reported by Tremonti et al. (2004). 
We used the B-band so that these results can be confronted with observational works 
where this band-width is often used.

\begin{figure}
\begin{center}
\vspace*{0.5cm}\resizebox{9.5cm}{!}{\includegraphics{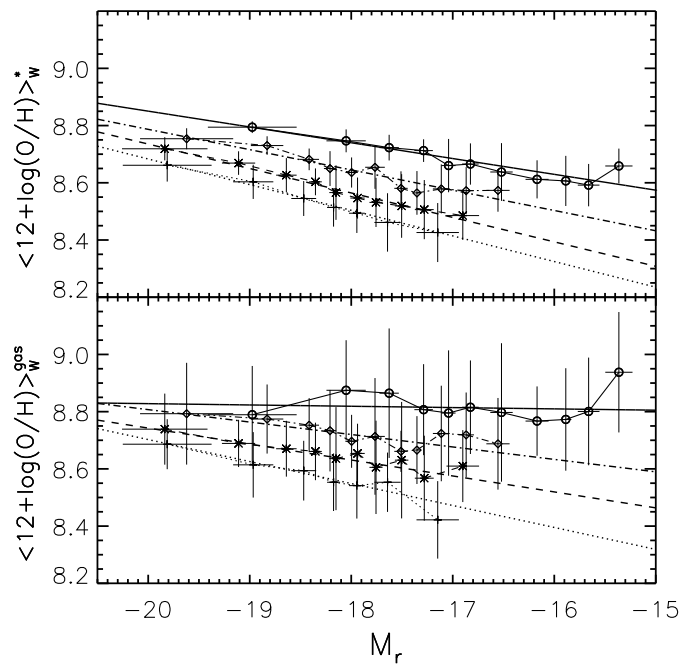}}\hspace*{-0.2cm}%
\end{center}
\caption[12+log(O/H) en funci\'on de $M_{\rm u}$, $M_{\rm g}$ y $M_{\rm r}$ a $z=0$]
{The LZR  for the stellar component (upper panel) and the 
gas-phase (lower panel) of 
simulated galaxies identified   at
$z=3$ (dotted line), $z=2$ (dashed line),
$z=1$ (dotted-dashed line) and $z=0$ (solid line). 
The error bars denote the rms deviations in bins of equal point number.}
\label{fig:LMR_z}
\end{figure}

Observational results  suggest an evolution in the zero point and slope of the LZR (Kobulnicky \& Koo 2000; 
Pettini et al. 2001;
Kobulnicky et al. 2003; 
Lilly, Carollo \& Stoctum 2003; 
Kobulnicky \& Kewley 2004; 
Maier et al. 2004; 
Shapley et al. 2004),
so that, at a given metallicity, on average, galaxies are brighter at higher redshifts.
To study the evolution of the simulated LZR, 
we estimated this relation at several redshifts and performed  linear regression fits through the data.
In Fig. ~\ref{fig:LMR_z}, we show the results of these
fits   for the mass-weighted  oxygen indicators for the stars (upper panel) and gas components (lower panel)
 at the  four redshifts of interest.
 We also include mean values estimated  in magnitude bins of equal point number. 
Table ~\ref{tab:LMR_z} summarizes the fitting parameters for all analysed redshifts.
As it can be appreciated from Fig. ~\ref{fig:LMR_z}, a linear regression is  a good representation of the data.
The results show that 
the simulated LZRs evolve in such a way that systems
become more metal-rich with time, at a given luminosity. The 
major changes are detected at the fainter end of the relation.
This evolution is not only true at a fix magnitude but also for individual galaxies which
tend to become fainter in ${\rm M_{\rm r}}$ with time.
Also, our simulations indicate that,
 at given metallicity, simulated galaxies  are, at least, $ 3$ magnitudes brighter 
at $z\sim 3$ than  systems at $z=0$ (see Shapley et al. 2004). 
We  estimate a mean overall  change of $\approx 0.15$ dex in the chemical content of the simulated galaxies from $z\sim 1$ (see Kobulnicky \& Kewley 2004).
Note that the LZR estimated from the gas-phase abundances have shallower slopes than those estimated from
the stellar components, mainly because of the high level of noise.

We stress the difficulties involved in the confrontation of  observations
with simulations owing  to different  uncertainties affecting the estimations of both  observed and simulated luminosities.
Taking into account the growing body of evidences that the fundamental relation is between metallicity and
stellar mass, hereafter, we will concentrate on the analysis of the MZR.

\begin{table}
\caption{\noindent Parameters for the linear regressions  of the local LZR: 12+log(O/H)= b + a  $\times \ M_{\rm B}$. 
The errors are shown in parenthesis}
\label{tab:LMR_z0}
\center
\begin{tabular}{lcc} \hline
\multicolumn{3}{c}{
12+log(O/H)=
b + a $\times \  M_{\rm B}$.} 
\\ \hline  \hline
Sample & b & a \\ \hline
Simulated stellar component & 7.74 (0.07) & -0.06 (0.01) \\ 
Simulated gas-phase & 8.73 (0.22) & -0.01 (0.01) \\
Tremonti et al. (2004) & 7.74 & -0.05 (0.01) \\ \hline
\end{tabular}
\end{table}

\begin{table*}
\caption{\noindent Parameters for the linear regressions of the simulated LZR relations: 12+log(O/H)= b + a  $\times \ M_{\rm r}$ as a function of redshift. }
\label{tab:LMR_z}
\center
\begin{tabular}{ccccc} \hline
& \multicolumn{2}{c}{Stellar Component}
& \multicolumn{2}{c}{Gas-phase}
\\ \hline  \hline
z & b & a & b & a\\ \hline
4.00 & 6.671  $\pm$ 0.239 &-0.098 $\pm$ 0.013  & 7.059  $\pm$ 0.305   & -0.077  $\pm$ 0.017\\   
3.50 & 7.024  $\pm$ 0.179 &-0.081 $\pm$ 0.010  & 7.210  $\pm$ 0.214   & -0.072  $\pm$ 0.012   \\
3.00 & 6.891  $\pm$ 0.130 &-0.090 $\pm$ 0.007  & 7.167  $\pm$ 0.202   & -0.077  $\pm$ 0.011  \\
2.50 & 6.750  $\pm$ 0.125 &-0.099 $\pm$ 0.007  & 7.113  $\pm$ 0.200   & -0.081  $\pm$ 0.011   \\
2.00 & 7.026  $\pm$ 0.117 &-0.085 $\pm$ 0.006  & 7.627  $\pm$ 0.185   & -0.056  $\pm$ 0.010  \\
1.80 & 7.259  $\pm$ 0.109 &-0.074 $\pm$ 0.006  & 7.695  $\pm$ 0.184   & -0.053  $\pm$ 0.010    \\
1.50 & 7.220  $\pm$ 0.108 &-0.077 $\pm$ 0.006  & 7.643  $\pm$ 0.179   & -0.057  $\pm$ 0.010   \\
1.30 & 7.414  $\pm$ 0.094 &-0.067 $\pm$ 0.005  & 8.136  $\pm$ 0.192   & -0.031  $\pm$ 0.011  \\
1.00 & 7.369  $\pm$ 0.092 &-0.071 $\pm$ 0.005  & 7.937  $\pm$ 0.184   & -0.044  $\pm$ 0.010   \\
0.80 & 7.366  $\pm$ 0.091 &-0.072 $\pm$ 0.005  & 8.227  $\pm$ 0.214   & -0.028  $\pm$ 0.012  \\
0.50 & 7.629  $\pm$ 0.076 &-0.059 $\pm$ 0.004  & 8.639  $\pm$ 0.214   & -0.008  $\pm$ 0.012    \\
0.12 & 7.651  $\pm$ 0.066 &-0.060 $\pm$ 0.004  & 8.604  $\pm$ 0.226   & -0.012  $\pm$ 0.013   \\
0.00 & 7.744  $\pm$ 0.070 &-0.055 $\pm$ 0.004  & 8.736  $\pm$ 0.218   & -0.005  $\pm$ 0.013    \\
\\ \hline
\end{tabular}
\end{table*}

\section{The Stellar Mass-Metallicity Relation}

New observational results are starting to provide information on the stellar mass and metallicity
of galaxies not only at $z=0$ but in the  intermediate  (Savaglio et al. 2005) and high redshift (Erb et al. 2006) 
universe. Despite the large observational difficulties to 
reliably estimate metallicities and stellar masses at high
redshift, a general picture, suggesting evolution, is starting to appear.
In fact, taking into account the detailed discussion presented by  Savaglio et al. (2005)
and Erb et al. (2006a),
a relative evolution of 0.1-0.15 dex at $z \sim 0.7$ and of $\sim 0.30$ dex at $z\sim 2$ has been determined.

\begin{figure}
\begin{center}
\vspace*{0.5cm}\resizebox{8.5cm}{!}{\includegraphics{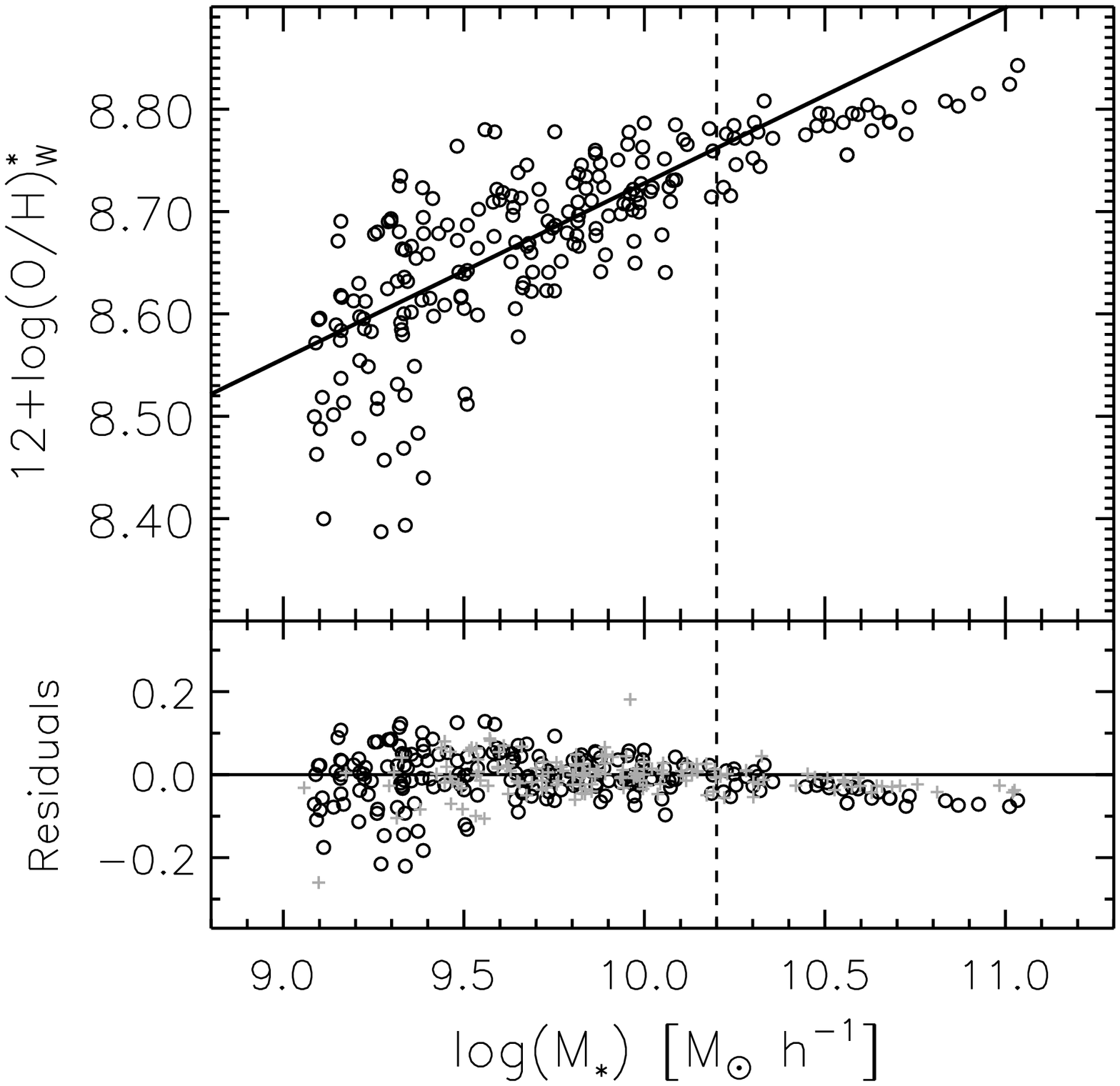}}\hspace*{-0.2cm}\\
\vspace*{0.5cm}\resizebox{8.5cm}{!}{\includegraphics{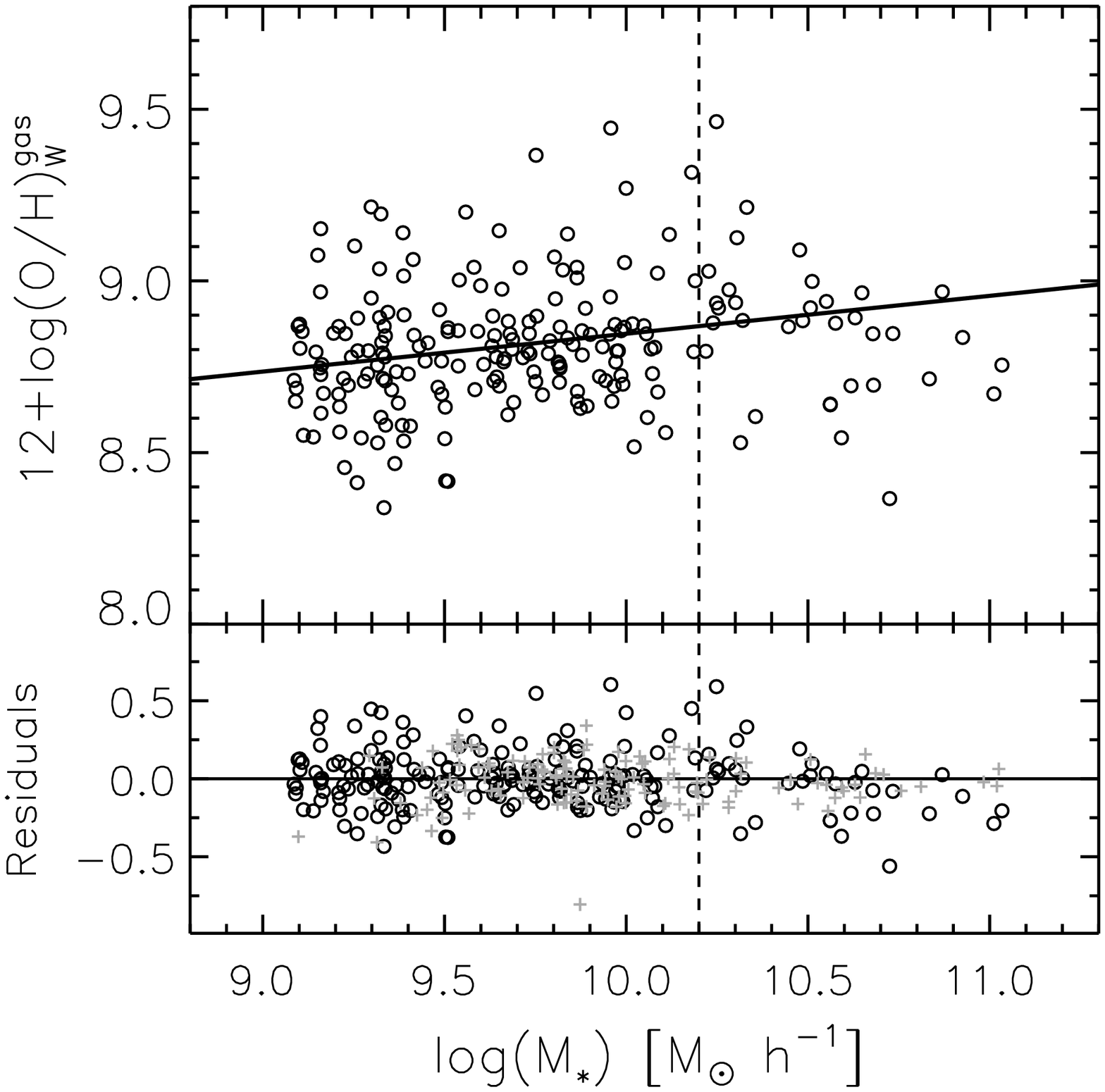}}\hspace*{-0.2cm}
\end{center}
\caption[12+log(O/H) en funci\'on de $M_{\rm u}$, $M_{\rm g}$ y $M_{\rm r}$ a $z=0$]
{Mass-Metallicity Relation at $z=0$
obtained for the stellar component (upper panel) and
gas-phase (lower panel) at $r_{\rm opt}$.   
Linear fits to the relations
are also shown (solid line) for
comparison. 
The dashed line indicates the characteristic
stellar mass where the residuals of the linear
fits (lower panels) to the stellar MZR
start to systematically depart 
from zero. In the lower panel the circles depicts S160
and the crosses S80.
}
\label{fig:MMR_z0}
\end{figure}

In the case of simulations, the estimation of the MZR is direct. 
The upper panel of Fig. \ref{fig:MMR_z0}  shows the
simulated local MZR using the stellar, mass-weighted O/H abundance as an indicator while
the lower panel displays the MZR estimated by using the gas abundances.
It is clear that the gas abundances have much larger dispersion. We carried out
linear regression fits to both distributions (solid lines).
As it can be appreciated from this figure, 
  a linear regression through the 
data  does not provide a good
representation of the simulated abundances for the whole range of stellar masses.
In small lower panels we displayed the residuals of these relations.
As it can be seen from these distributions,
 for stellar masses larger than $ M_c \approx 10^{10.2} M_{\sun} h^{-1}$, the residuals
become systematically negative, indicating a flattening of  the slope 
for   systems with larger stellar masses. A slight higher $ M_c \ (\approx 10^{10.4} M_{\sun} h^{-1})$
is suggested from the abundances in gas phase distribution. However, due to the high dispersion
present in the gas-phase data, we take  $ M_c \approx 10^{10.2} M_{\sun} h^{-1}$ as the reference mass.
We have also included the same estimations for the simulated galaxies in S80 (gray crosses) in order
to assess the effects of resolution in the determination of $M_c$. As it can be seen
both simulations (which differ by approximately an order of magnitude in mass  resolution) yield 
similar results.

Interestingly, this  stellar mass $ M_c$ is not far from the characteristic mass $\approx 10^{10.5} M_{\sun} h^{-1}$  determined by  
Kauffmann et al. (2004), which is claimed by these authors
 to separate  spheroid-dominated galaxies from  spiral-dominated ones.
T04 also determined that their MZR changes curvature at approximately the same stellar mass.
Hence, the fact that this characteristic mass is also found in our simulations suggests that 
 the way in which building blocks are assembled might play a relevant role in its determination.

In order to make a more direct comparison of our data with
observations (see also following section), we also estimated
$M_c$ by computing the mean oxygen abundance within 
$r_{\rm ap}$ and  the residuals
of the linear fits to the resulting MZRs.  We obtained
that limiting our analysis to the central region of galaxies
shifts the caracteristic $M_c$ of the stellar MZR to $\approx 10^{10.3} M_{\sun} h^{-1}$
and that of the gas phase one to $\approx 10^{10.5} M_{\sun} h^{-1}$,
approaching the observed value. However, it is important to note
that if Kauffmann et al. (2004) or T04 had assumed a Salpeter
IMF, their characteristic mass would have been about 0.2 dex higher and
hence even higher than the one obtained from simulations.

\begin{figure*}
\begin{center}
\vspace*{0.5cm}\resizebox{16.5cm}{!}{\includegraphics{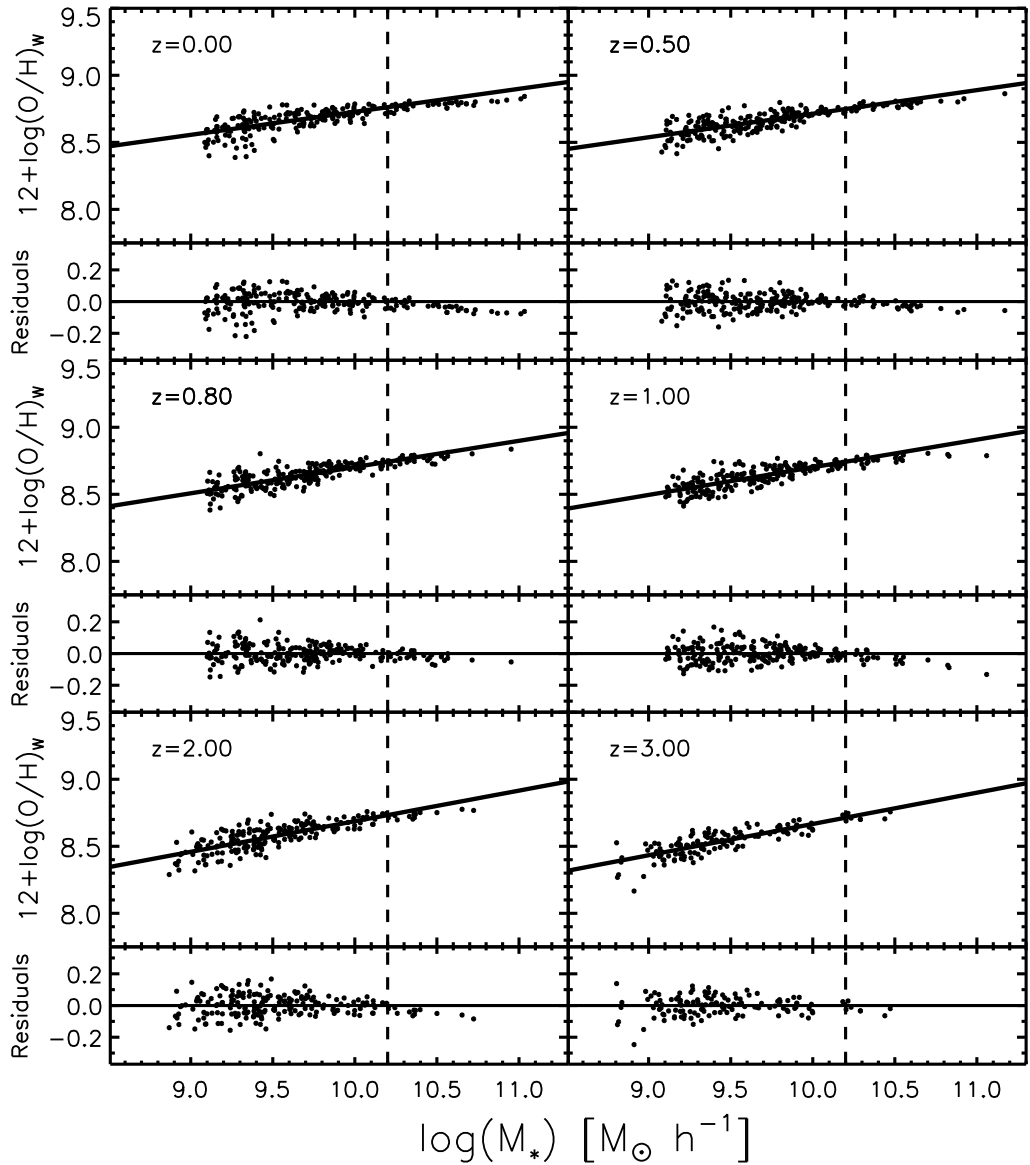}}\hspace*{-0.2cm}
\end{center}
\caption
{The MZR for simulated galaxies at different redshifts
and the residuals (small boxes)  estimated with respect to  
 linear regression fits (solid lines).
The dashed lines indicate the characteristic stellar
mass where the residuals  
 start
to systematically depart from zero. 
}
\label{fig:MMR_z}
\end{figure*}

In order to investigate the origin of the MZR and the $M_c$, 
we  estimated  this relation as a function of redshift.
In the case of  $M_c$, we used the MZR estimated from the stellar abundances because of their  lower dispersion with
respect to those obtained from the gas abundances and measured at $r_{\rm opt}$. These abundances provide a better
estimation of the characteristics of the galaxies to study their chemical evolution.
From Fig. \ref{fig:MMR_z}, we observe that the MZR exhibits
the same general patterns from $z=0$ to
$z=3$, but with a displacement toward
larger abundances for decreasing redshift. 
In the small panels of  Fig. \ref{fig:MMR_z}, we show the residuals of each relation with respect to
the corresponding linear regression fit. As we can appreciate, at all analysed redshifts, we find the same behaviour:
the residuals became systematically negative for $M_* > M_c$.

The characteristic mass remains almost unchanged 
from $z=3$,  with an evolution of its 
abundance by only 0.05 dex in the same 
redshift range. 
As discussed by Tissera et al. (2005), the major variations in the chemical content
 are found for systems 
 with  $M_* < M_{\rm c}$  which,  on average,  increase their  abundances
 by $\approx  0.20$ dex from $z \sim 3$ to $z \sim 0$.
Systems with $M_* > M_{\rm c}$    show an increase in the level of 
enrichment of     $\approx  0.05$ dex in the same redshift range.
Note that even at $z> 2$, galaxies with $M_* > M_c$ can have stellar populations with mean
solar abundances.

From  Fig. \ref{fig:MMR_z} we can also appreciate that, at a given stellar mass,  the dispersion
in oxygen abundances increases for higher redshift, and that the dispersion is larger for smaller
stellar mass systems. 
The different evolution found for simulated galaxies of different masses and the dependence
of the dispersion on stellar mass are produced by their
different evolutionary paths which establish the way the systems are assembled and how
the gas is consumed into stars in these simulations where SN feedback has not  been included yet.

\subsection{Comparison with observations}

As discussed by Tissera et al. (2005), the level of evolution obtained in the metallicity content of
the simulated galaxies in models without feedback is weaker than that suggested by recent observations.
These authors chose the mean O/H abundances in the simulated stellar populations to estimate the MZR
because of the high dispersion present in the gas abundances.
In  order to improve the comparison with observations, in this Section we will use the gas abundances
estimated at the aperture radius ($r_{\rm apt}$) to assess the evolution predicted by our model.
This is because the SDSS explored only the
central regions of galaxies  enclosing approximately $1/3$ of the luminosity of the galaxies.

\begin{figure}
\begin{center}
\vspace*{0.5cm}\resizebox{8.5cm}{!}{\includegraphics{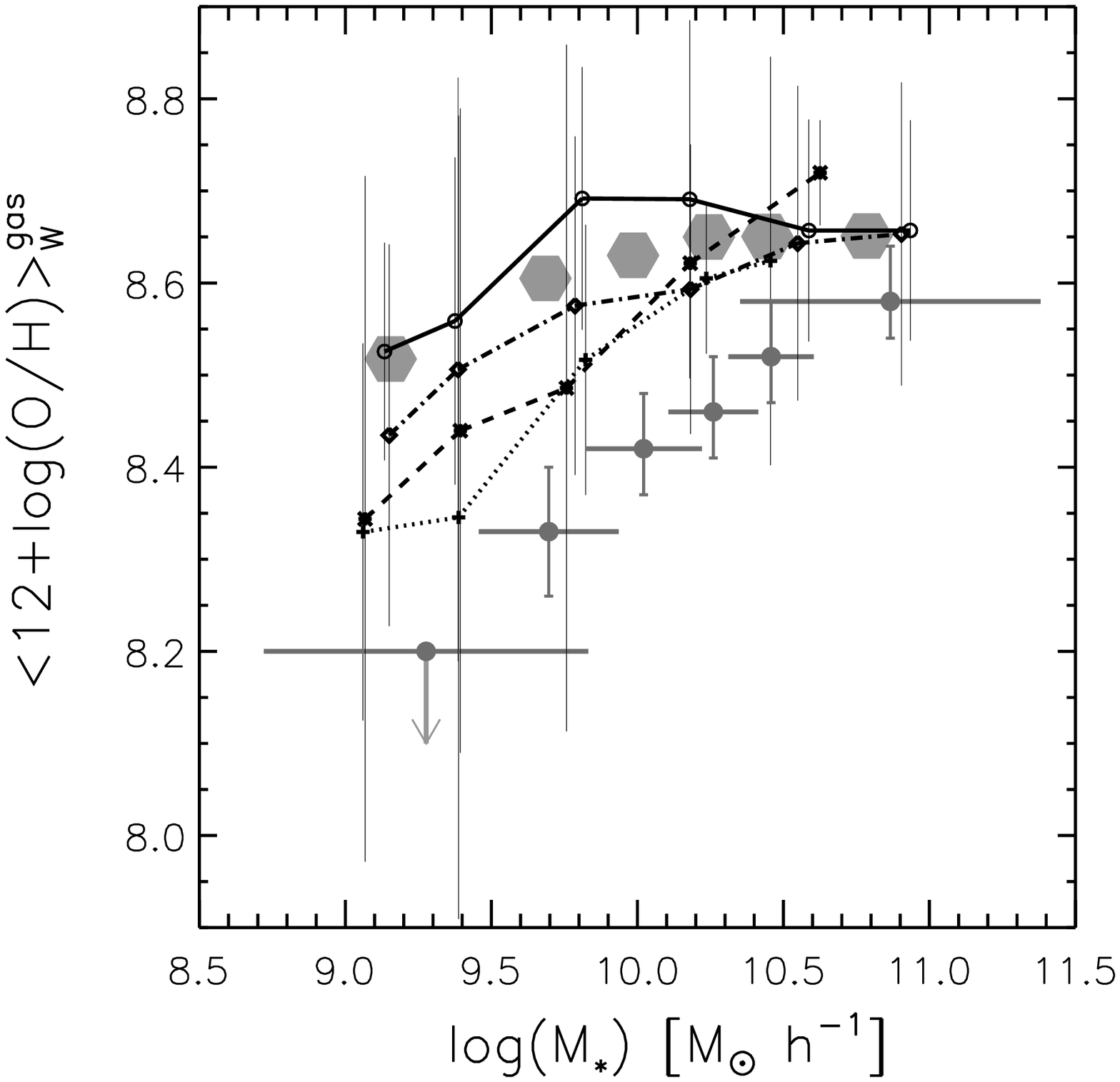}}\hspace*{-0.2cm}
\end{center}
\caption
{
The gas-phase MZR  
for simulated galaxies at different redshifts (see Fig. ~\ref{fig:LMR_z} for line code). 
The oxygen abundances are calculated at $r_{\rm ap}$.
Filled circles denote the observed MZR at $z \sim 2$ by Erb et al. (2006a) while
the large exagons correspond to the observed MZR at $z=0$ by  Tremonti et al. (2004)
as recalculated by Erb et al. (2005a).
 Note that we have displaced the simulated MZRs by -0.26 dex in 
order to fit the gas-phase abundances of Tremonti et al. (2004)
at low stellar masses and that this displacement roughly agrees with that estimated
by Erb et al. (2005a) for the T04 relation.
}
\label{comp}
\end{figure}

In Fig. ~\ref{comp} we plot the simulated MZR from $z=3$ to $z=0$ (see also Table 3), 
the observations of
Erb et al. (2006a) and the observed MZR of T04 as obtained by Erb et al. (2006a) by using
their own metallicity estimator.
Erb et al. (2006a) performed a detailed comparison between both sets of observations, concluding
that an evolution of 0.30 dex from $z\approx 2$ to $z=0$ could be reliably claimed. 
The simulated MZR has been shifted by -0.26 dex in order to fit the observational
MZR of T04 for low mass systems, as the metallicity indicator employed by Erb et al. (2006a) 
saturates at higher 
metallicities and can not be trusted, according to those authors. Note that the displacement
applied to the simulated MZR at $z \sim 0$ is consistent with the global displacement
estimated by Erb et al. (2005a) for the MZR of T04. Hence, at $z\sim 0$ our simulated
MZR calculated at $r_{\rm apt}$ agrees with the that of T04 without requiring any extra correction.

As it can be appreciated, in the simulations, 
we found  a lower level of evolution in comparison with observations, with a change of $\approx 0.05$ dex for the
high stellar mass end and $\approx 0.1-0.15$ dex for the low stellar  mass end. 
From this figure, we can see that a better agreement
between simulations and observations seems to require SN winds in order 
to decrease the metallicity content not only for low stellar mass systems but also for large stellar masses.  
This last issue is still controversial because, as reported by Erb et al. (2006), their metallicity indicator could
be saturated for high metallicities and hence, the authors do not assure that the level of evolution they reach for
larger masses is, actually, correct. 
Also, in the simulations and observations, different
IMFs have been adopted, fact that introduces an uncertainty in the determination of the 
stellar mass of up a factor of 2, among them.

\begin{table*}
\caption{Mean simulated gas-phase O/H abundances estimated within $r_{\rm ap}$ for different stellar mass bins at 
$z=3.0,2.0,1.0,0.0$.  The number N of galaxies identified at each
redshift is also shown.}
\label{tab:mean_MZR_z}
\center
\begin{tabular}{lcccccc} \hline
\multicolumn{1}{c}{${\rm M}_{*} [{\rm M}_{\sun}  h^{-1}]$} &
\multicolumn{6}{c}{$<$12+log(O/H)$>$}
\\ \hline \hline 
$\log({\rm M}_{*})<9.20$   & 8.59 $\pm$ 0.20 & 8.60 $\pm$ 0.37 & 8.69 $\pm$ 0.21 & 8.79 $\pm$ 0.12 & 8.82 $\pm$ 0.17 & 8.79 $\pm$ 0.12\\
$9.20 \le \log({\rm M}_{*})<9.60$   & 8.61 $\pm$ 0.44 & 8.70 $\pm$ 0.35 & 8.77 $\pm$ 0.32 & 8.82 $\pm$ 0.18 & 8.79 $\pm$ 0.47 & 8.82 $\pm$ 0.18\\
$9.60 \le \log({\rm M}_{*})<10.00$  & 8.78 $\pm$ 0.15 & 8.75 $\pm$ 0.37 & 8.84 $\pm$ 0.18 & 8.95 $\pm$ 0.14&8.91 $\pm$ 0.15& 8.95$\pm$ 0.14\\
$10.00 \le \log({\rm M}_{*})<10.40$ & 8.87 $\pm$ 0.08 & 8.88 $\pm$ 0.10 & 8.85 $\pm$ 0.16 & 8.95 $\pm$ 0.19&8.89 $\pm$ 0.14 & 8.95 $\pm$ 0.19\\
$10.40 \le \log({\rm M}_{*})<10.80$  & 8.88 $\pm$ 0.22 & 8.98 $\pm$ 0.06 & 8.90 $\pm$ 0.17 & 8.92 $\pm$ 0.12&8.89 $\pm$ 0.13& 8.92 $\pm$ 0.12\\
$10.80 \le \log({\rm M}_{*})$0 & ------ & ------ & 8.91 $\pm$ 0.16 & 8.92 $\pm$ 0.12&8.90 $\pm$0.12 & 8.92 $\pm$ 0.12\\
\hline \hline 
 z    &   3.00  &     2.00     &   1.00   &     0.80     &   0.50    &     0.00\\
 N    & 137    &  202    &  241    &  239   &   239   &   227    

\\ \hline  

\end{tabular}
\end{table*}

\begin{table*}
\caption{Mean simulated stellar O/H abundances estimated within $r_{\rm ap}$ for different stellar mass bins at
$z=3.0,2.0,1.0,0.8,0.5,0.0$.  The number N of galaxies identified at each
redshift is also shown.}
\label{tab:mean_MZR_z}
\center
\begin{tabular}{lcccccc} \hline
\multicolumn{1}{c}{${\rm M}_{*} [{\rm M}_{\sun}  h^{-1}]$} &
\multicolumn{6}{c}{$<$12+log(O/H)$>$}
\\ \hline \hline
$\log({\rm M}_{*})<9.20$  & 8.40 $\pm$ 0.34 & 8.46 $\pm$ 0.26 & 8.52 $\pm$ 0.15 & 8.56 $\pm$ 0.10 & 8.58 $\pm$ 0.08 & 8.62 $\pm$ 0.08\\
$9.20 \le \log({\rm M}_{*})<9.60$  & 8.49 $\pm$ 0.29 & 8.55 $\pm$ 0.29 & 8.62 $\pm$ 0.13 & 8.59 $\pm$ 0.28 & 8.66 $\pm$ 0.11 & 8.66 $\pm$ 0.14\\
$9.60 \le \log({\rm M}_{*})<10.00$  & 8.65 $\pm$ 0.10 & 8.65 $\pm$ 0.14 & 8.71 $\pm$ 0.07 & 8.69 $\pm$ 0.12 & 8.72 $\pm$ 0.07 & 8.75 $\pm$ 0.10\\
$10.00 \le \log({\rm M}_{*})<10.40$  & 8.77 $\pm$ 0.02 & 8.76 $\pm$ 0.06 & 8.78 $\pm$ 0.04 & 8.76 $\pm$ 0.08 & 8.80 $\pm$ 0.03 & 8.80 $\pm$ 0.06\\
$10.40 \le \log({\rm M}_{*})<10.80$ 0 & 8.79 $\pm$ 0.01 & 8.82 $\pm$ 0.03 & 8.81 $\pm$ 0.03 & 8.76 $\pm$ 0.12 & 8.83 $\pm$ 0.02 & 8.83 $\pm$ 0.02\\
$10.80 \le \log({\rm M}_{*})$  & ------- & ------- & 8.82 $\pm$ 0.04 & 8.89 $\pm$ 0.00 & 8.89 $\pm$ 0.04 & 8.87 $\pm$ 0.03\\
\hline \hline
 z    &   3.00  &     2.00     &   1.00   &     0.80     &   0.50    &     0.00\\
  N    & 137    &  202    &  241    &  239   &   239   &   227

  \\ \hline

  \end{tabular}
  \end{table*}

\section{Discussion}
In order to understand the physical meaning of the evolution of the
MZR and of the
characteristic mass, $M_{\rm c}$,
 we have analysed the merger trees of the simulated systems at $z=0$
and the variation of the astrophysical properties along the evolutionary paths.
\subsection{Does $M_c$ segregate two different types of galaxies?}

In  previous sections, we investigated the relation between the metal content of the simulated galaxies
in a $\Lambda$-CDM scenario and their dynamical and astrophysical properties finding that the chemical
enrichment of systems increase with luminosity and  stellar mass.
When estimating the MZR as a function of redshift, we detected a characteristic stellar mass ($M_{\rm c} \sim 10^{10.2}
{\rm M_{\odot}} h^{-1}$)
 at which
the relation changes curvature toward a flatter slope.
 This $M_{\rm c}$ seems to be approximately constant with redshift, reaching 
solar abundance at $z \approx 0$ but with an overall chemical evolution of only  $\approx 0.05$ dex since $z\approx 3$. 

Interestingly, the simulated $M_{\rm c}$  agrees with a similar motivated stellar mass determined by Tremonti et al. (2003)
in an observational study of the MZR for galaxies in the SDSS-DR2.
It is also worth noting the consistency of $M_{\rm c}$ with
the characteristic mass derived by Kauffmann et al. (2004) 
when analysing the astrophysical properties of galaxies
in SDSS. Note that the simulated $log M_{\rm c}$ can vary in the range $10^{0.5}-10^{0.2}$
 depending on the region
where abundances are estimated (i.e. $r_{\rm opt}$ and $r_{\rm ap}$ and on the adopted metallicity estimator
(i.e. gas or stellar metallicities). 
Kauffmann et al. (2004) found that galaxies with large stellar masses tend to be dominated by old stars, in 
spheroid-type systems with red colours, while smaller systems determine a different galaxy population with their
astrophysical properties corresponding to younger, blue, disk-dominated galaxies.
In this section, we assess if  $M_{\rm c}$ can separate the simulated galaxies into 
two different populations.

\begin{figure*}
\begin{center}
\vspace*{0.5cm}\resizebox{14.5cm}{!}{\includegraphics{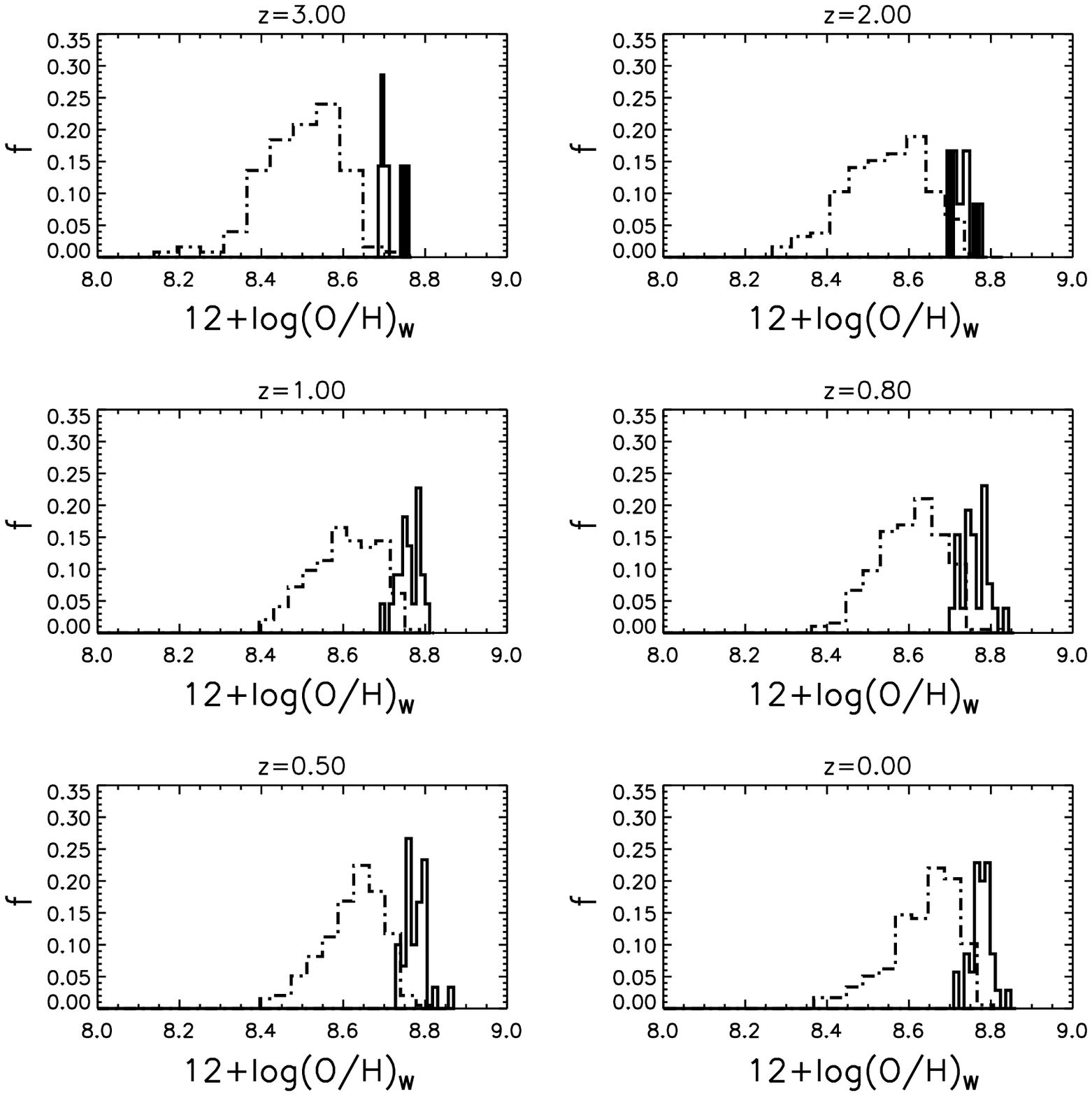}}\hspace*{-0.2cm}%
\end{center}
\caption
{Mass-weighted oxygen abundance distributions of the stellar populations for galaxy-like  
systems with $M_* < M_c $ (dotted-dashed lines) and $M_* > M_c $ (solid lines) at different redshifts.}
\label{fig:hist_OH}
\end{figure*}

In Fig.~\ref{fig:hist_OH} we show the distributions of O/H abundances of the stellar populations in   the simulated
galaxies separated into two groups depending on their stellar mass. Massive systems ($M_* > M_{\rm c}$)
have mean abundances comparable to solar levels
from $z\sim 3$.
 The distributions
are quite narrow (with a width of $\approx 0.10 $ dex) indicating that the composition of the stellar populations in  these systems are very similar
even at  $z\sim 3$, experiencing   no significant
changes in the analysed redshift range.
 Conversely, simulated galaxies with  $M_* < M_{\rm c}$  show a broader metallicity distribution which
evolves to   higher level of enrichment  with redshift.

\begin{figure*}
\begin{center}
\vspace*{0.5cm}\resizebox{14.5cm}{!}{\includegraphics{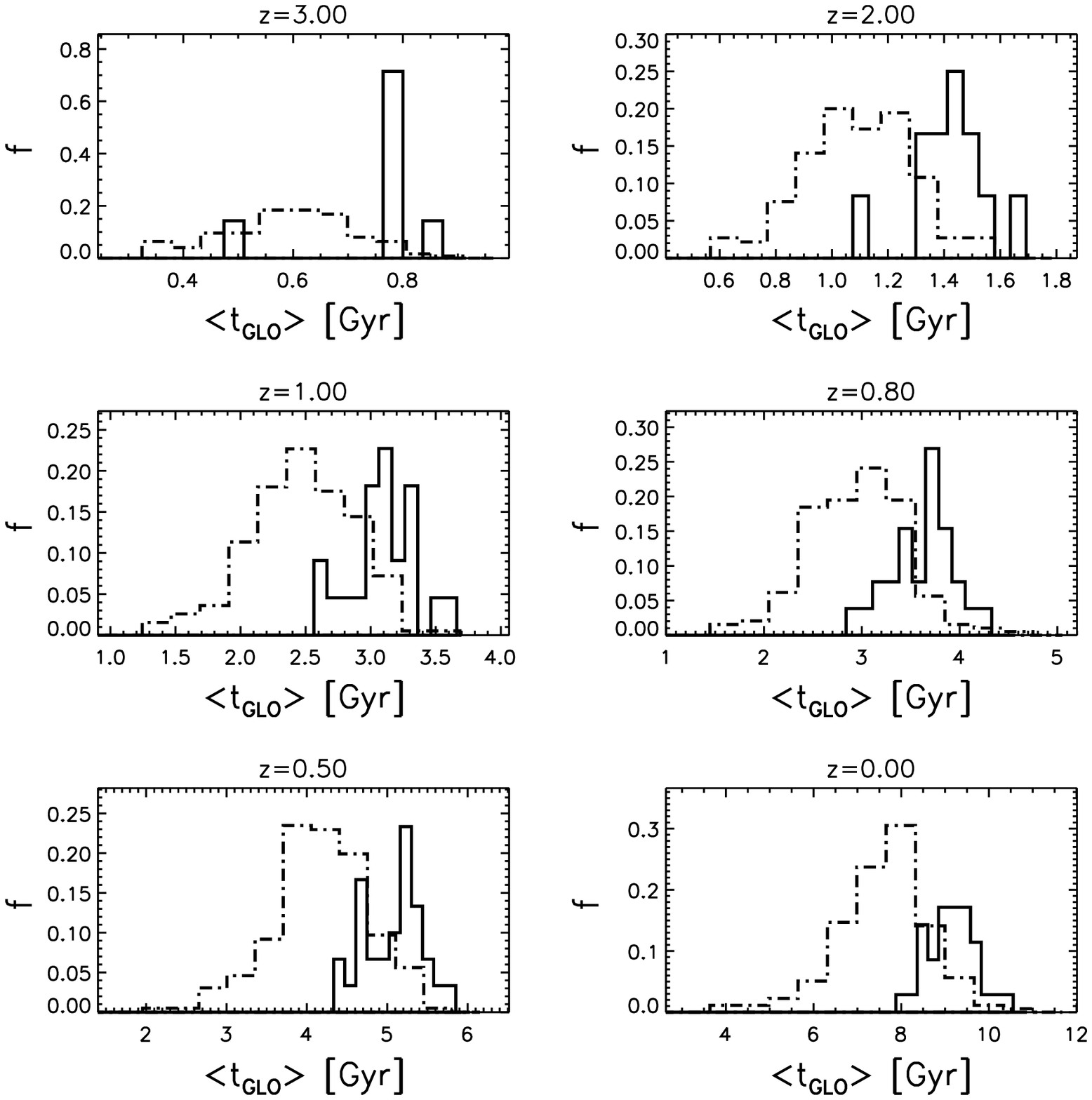}}\hspace*{-0.2cm}%
\end{center}
\caption
{Distributions of the mean  ages of stellar population in  simulated
galaxies  with $M_* < M_c $ (dotted-dashed lines) and
with $M_* > M_c $ (solid lines) at different redshifts.
}
\label{fig:hist_tglo}
\end{figure*}

The fact that massive systems are dominated by old stellar populations can be confirmed by inspecting
Fig. ~\ref{fig:hist_tglo} where we show a similar plot to that of  Fig.~\ref{fig:hist_OH} but as
a function  of the
mean stellar age of the simulated galaxies. As it can be appreciated, massive systems have always 
mean older populations than the rest of the simulated systems. Again, smaller systems show a broader distribution with 
some of them  having mean ages comparable to those of massive ones. However, the bulk of the stellar
population in systems with  $M_* < M_{\rm c}$  is clearly younger than their massive counterparts.

\begin{figure*}
\begin{center}
\vspace*{0.5cm}\resizebox{14.5cm}{!}{\includegraphics{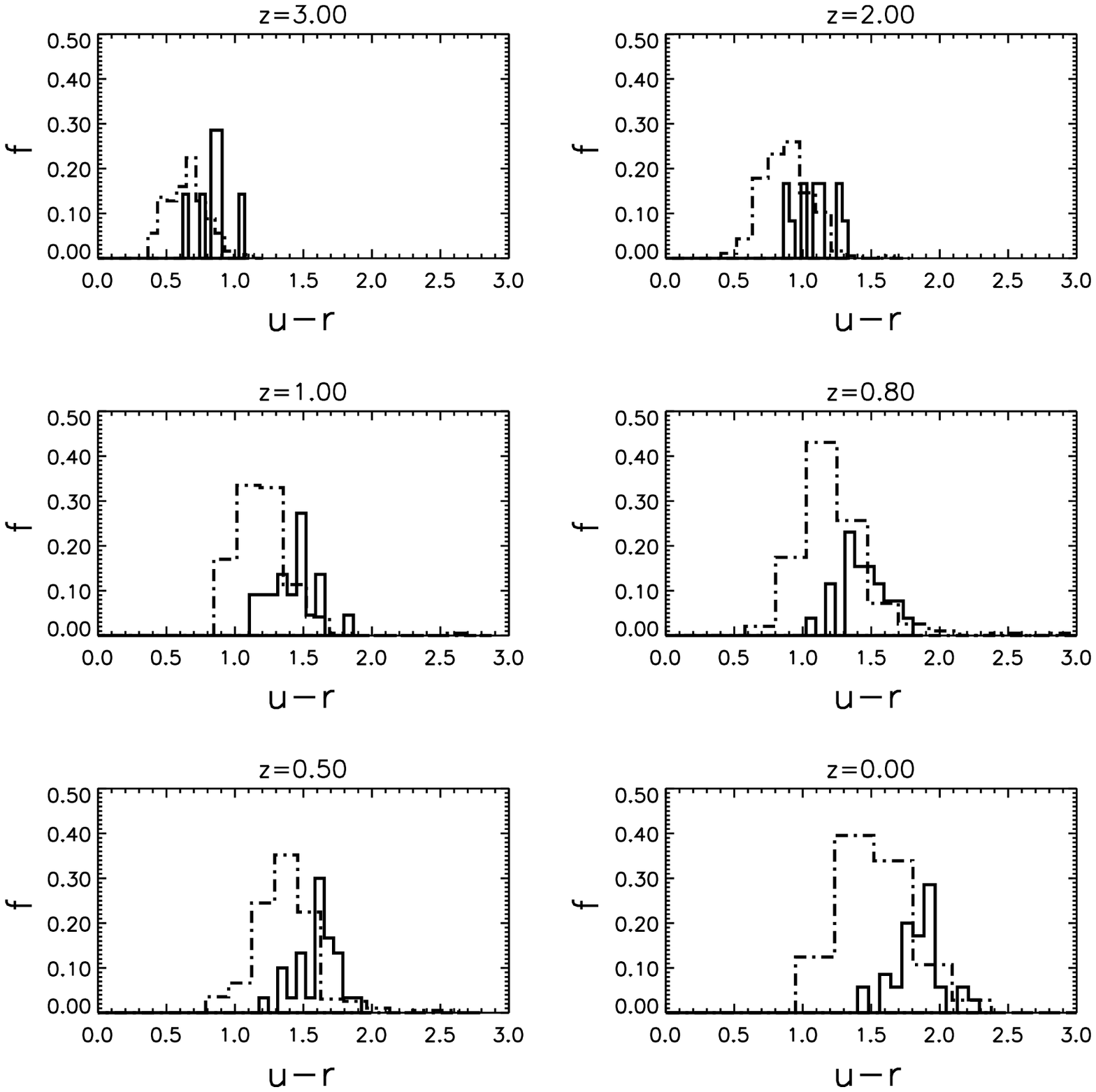}}\hspace*{-0.2cm}%
\end{center}
\caption
{$u-r$ colour distributions in  simulated
galaxies  with $M_* < M_c $ (dotted-dashed lines) and
with $M_* > M_c $ (solid lines) at the  redshifts of interest.
}
\label{colours}
\end{figure*}

These figures show that  $M_{\rm c}$  segregates two distinctive  galaxy populations with different histories of formation.
The simulated massive galaxies are dominated, on average, by older and more metal-rich stellar populations than
smaller systems. 
By combining metallicity and age information with the population models of BC06,
we estimated colours as a function of redshift. As we can observe from Fig. ~\ref{colours},
$M_c $ also segregates galaxies into two clear colour distributions, with massive
systems in the red tail of the distribution, even from $z \sim 3$. 
This finding is in agreement whit recent observational results obtained from the VIMOS-VLT Deep Survey
of galaxies at $z\sim2$ (Franzetti et al. 2006).
In these simulations, blue and red colours which
produce the bimodal distribution
are determined by the star formation history of the systems
which establishes 
the relative fraction of old to new stars and their metallicities.
We note that the bimodal
distribution at $z=0$ is not in full agreement with observations (Balogh et al. 2004), principally 
because we are lacking a fraction of very red systems ($u-r>2$).
There are two possible reasons for this. 
First, it is possible that the action of SN feedback could contribute to modulate a bimodal colour distribution
in full agreement with that observed at $z=0$. 
Secondly, the small volume simulated 
in this work could prevent us from reproducing the complete range of observed colours.

These trends show that in these simulations
$M_{\rm c}$ segregates two different galaxy populations
in a similar sense to the trend
reported by Kauffmann et al. (2004).
Our results also indicate that
 $M_{\rm c}$ remains as a characteristic mass up to $z\approx 3$.
Our findings suggest that the origin of  $M_{\rm c}$ can be directly linked to the way the structure is
built up in a hierarchical Universe.
Supernova feedback is expected to play a relevant role in the regulation of the star formation activity.
However, this process might not  significantly  affect the  $M_{\rm c}$ 
as we will discuss in the next 
Section.

\subsection{The hierarchical building up of the structure}

From the   analysis of the merger trees, we find that systems with
$M_* <M_{\rm c}$
transform their gas content into stars in a more passive
fashion or via gas-rich mergers,  setting
a steeper correlation between stellar mass and  metallicity (Tissera et al. 2005). In this case, 
the fraction of new born stars in  merger events is high enough so that
the mean metallicity of the remanent systems is affected by the new stellar contribution.

Galaxies with masses larger than
  $M_{\rm c}$ are preferentially assembled by  merger events which involve stellar
dominated systems, with small leftover gas .
In these cases,  the mergers  produce  systems with  final stellar masses
equal to the sum of the old-dominating stars   plus some new born ones,
while its overall mean abundance remains at the same level of enrichment.
This situation
occurs because, in this case, the merging systems have already transformed most of their gas into
stars so that there is no  fuel for an important starburst during the merger.
It could be also possible that a large system merges with a smaller less-enriched one
which can feed  new star formation activity but with lower level of enrichment.
Both scenarios have the same flattening effect on the slope of
the mass-metallicity relation in
systems within this range of masses.
We also estimated the fraction of gas in the simulated galaxies as a function of their stellar mass from 
$z\approx 3$. From this analysis, we  found that systems with  
 $M_* \sim M_{\rm c}$ have, on average, $15 \%$ of leftover gas (Fig.~\ref{gasfrac}) at all analysed redshifts.

\begin{figure}
\begin{center}
\vspace*{0.5cm}\resizebox{8.5cm}{!}{\includegraphics{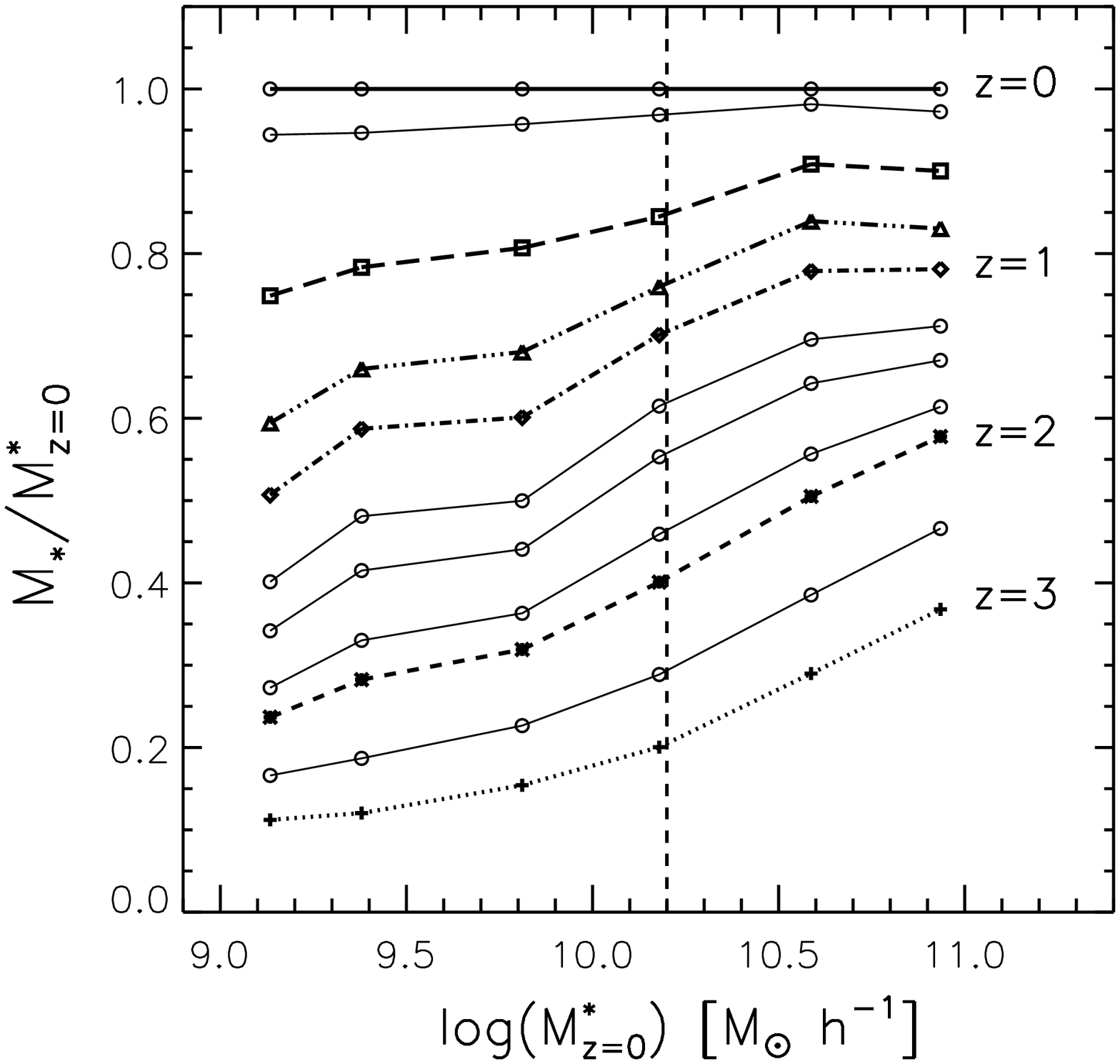}}\hspace*{-0.2cm}
\end{center}
\caption
{Fraction of total stellar mass in simulated galaxies at $z=0$
formed  at different redshifts. From bottom
to top the lines correspond to 
$z=3.0,2.5,2.0,1.8,1.5,1.3,1.0,0.8,0.5,0.1,0.0$.}
\label{fig:Mform_z}
\end{figure}

How stars in systems of different masses are formed        can be 
seen in Fig. ~\ref{fig:Mform_z} where we show the fraction of stellar mass  in the simulated galaxies identified at
$z=0$ which formed at different $z$. As it can be appreciated from this plot, simulated galaxies with $M_* > M_c$ have $50$ per cent of their
stars formed at $z \ge 1.5$ with less than 30 per cent appearing after $z \approx 1$.
In the case of simulated galaxies with  $M_* < M_c$, they have 50 percent of the stars formed after $ z\approx 1.3$.
Note that  these stars would probably belong to different substructures which
then merged to build up our simulated galaxies at $z=0$.
 In other words, our results do not imply that, a given simulated galaxy at $z=0$ was formed at
certain redshift, but that a given  fraction of its stars was born at that time.
 These stars could have belonged to different
substructures by the redshift of reference. The ensemble of the final systems can occur at lower redshift resulting
in a natural ``downsizing'' scenario (see also De Lucia et al. 2006; Neistein et al. 2006).

\subsection{The missing process: SN energy feedback}

As we mentioned before,
we have not included an effective SN energy feedback treatment in our simulations,
which is expected to produce powerful outflows, affecting
the star formation process and the metal production and mixing.
Taking into account theoretical results, the effects of SN feedback depend
strongly on the potential well of the systems and  is expected to
affect mainly systems with circular velocities  smaller than $100 \kms$ (e.g. Larson 1974).
The simulations analysed in the paper do not include a strong SN energy feedback that can
actually produce an impact on the dynamical evolution and star formation activity of the systems
(Scannapieco et al. 2006).
Nevertheless, we can assess the effects we would expect.

\begin{figure}
\begin{center}
\vspace*{0.5cm}\resizebox{8.5cm}{!}{\includegraphics{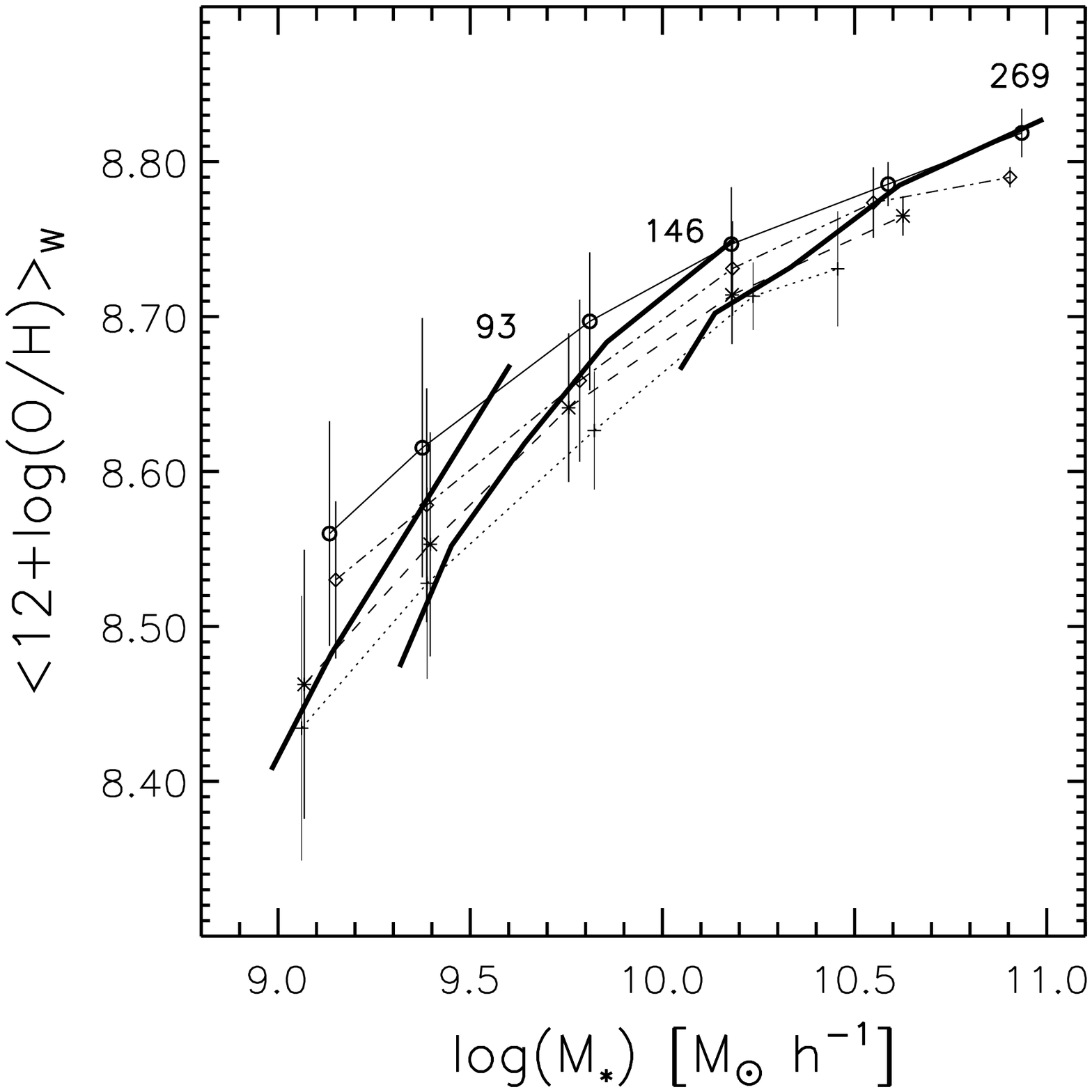}}\hspace*{-0.2cm}\\
\vspace*{0.5cm}\resizebox{8.5cm}{!}{\includegraphics{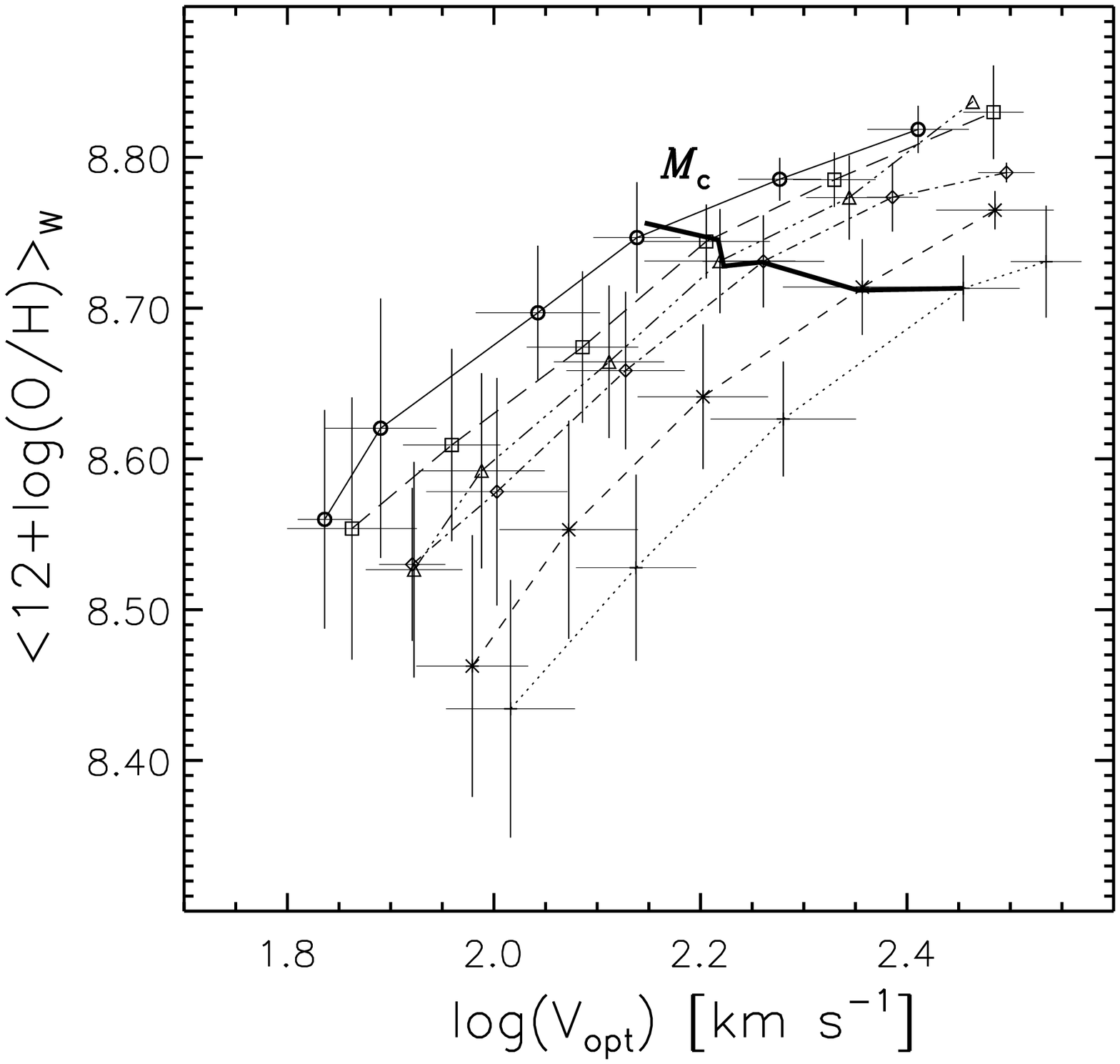}}\hspace*{-0.2cm}%
\end{center}
\caption
{Mean MZR (upper panel) and VZR (lower panel) for galactic systems at
$z=3$ (dotted line),
$z=2$ (dashed line),
$z=1$ (dotted-dashed line)
and $z=0$ (solid line).
Upper panel: The thick solid lines represent
curves of constant optical velocities
(93  ${\rm km \ s^{-1}}$, 146  ${\rm km \ s^{-1}}$ and 269 ${\rm km \ s^{-1}}$). Lower panel:
the thick solid line depicts the line of constant $M_c$.
Results at $z=0.5$ (long-dashed line) and $z=0.8$ (triple-dotted-dashed line)
are also shown.
The error
bars correspond to standard deviations estimated in bins of $\sim 0.4$ dex
in log $M_*$. Note that the VZRs for $z=3$ and $z=2$ do not have the bin corresponding
to the largest stellar mass interval owing to the lack of a statistical  number of systems
with high stellar mass at very high redshift in our simulated volume.
}
\label{fig:MMR_V_z0}
\end{figure}

In  Fig. \ref{fig:MMR_V_z0} we show the  MZR  (upper panel) and velocity-metallicity relation 
(VZR; lower panel) from $z=3$ to $z=0$. 
The optical velocity has been estimated as $V_{\rm opt}= (G M (r < r_{\rm opt}) / r_{\rm opt})^{0.5}$ where 
$M (r < r_{\rm opt})$ is the total mass within the optical radius, $r_{\rm opt}$.
Hence,  $V_{\rm opt}$ is a measure of the potential well of the system.

From this figure it can be appreciated that
the chemical content of the simulated galaxies tends
to increase with  optical velocities.  
Fast rotating systems increase their
chemical abundances by $\sim 0.10$ dex 
from $z=3$ to $z=0$, while
at lower velocities the variations are 
of $\sim 0.25$ dex. 
From these figures, we can see how, on average, at a certain stellar mass, systems become slower rotators
and  more chemical enriched with decreasing redshift. At a given circular velocity, the evolution
in metallicity is larger since,  at lower redshift,  systems of larger stellar masses contribute to
this given circular velocity. 
The VZR shows a higher level of  evolution with redshift than the MZR
as  a consequence of the combination of metallicity enrichment and cosmology.
The increase of the mean 
density of the universe at higher redshift produces  that,
at a given stellar mass, systems need to be more concentrated
in order to get gravitationally bounded and separate from  the general expansion.
Then, for a given stellar mass the virialization of the system  occurs at   higher circular velocities
for  higher redshift.

Assuming that SN energy feedback will affect more strongly slow rotating systems (i.e. Larson 1974; 
Dekel \& Silk 1986), we expect that
the  $M_{\rm c}$ may not be strongly modified by this process. This is because, 
as seen in  Fig.~\ref{fig:MMR_V_z0}, the optical velocity corresponding to  $M_{\rm c}$ 
varies from around $300 \kms$ at $z=3$ to $ 140 \kms$ at $z=0$.
However, because the hierarchical aggregation of the structure, large systems might be affected
through the strong action of SN energy feedback in the substructure that merged to form them. 

It is clear that since the strongest effects of SN 
energy feedback are expected to
take place in the small mass range,
the excess of metals in the MZR shown  by Tissera et al. (2005) could be solved by this process.
 The action of strong SN   feedback
would produce  the ejection of  part of the enriched material out of the systems and the  decrease of
the  general  level of enrichment making the slope of the simulated MZR for $M_* < M_c$ steeper as we expect from
the observational relation.

\begin{figure}
\begin{center}
\vspace*{0.5cm}\resizebox{8.5cm}{!}{\includegraphics{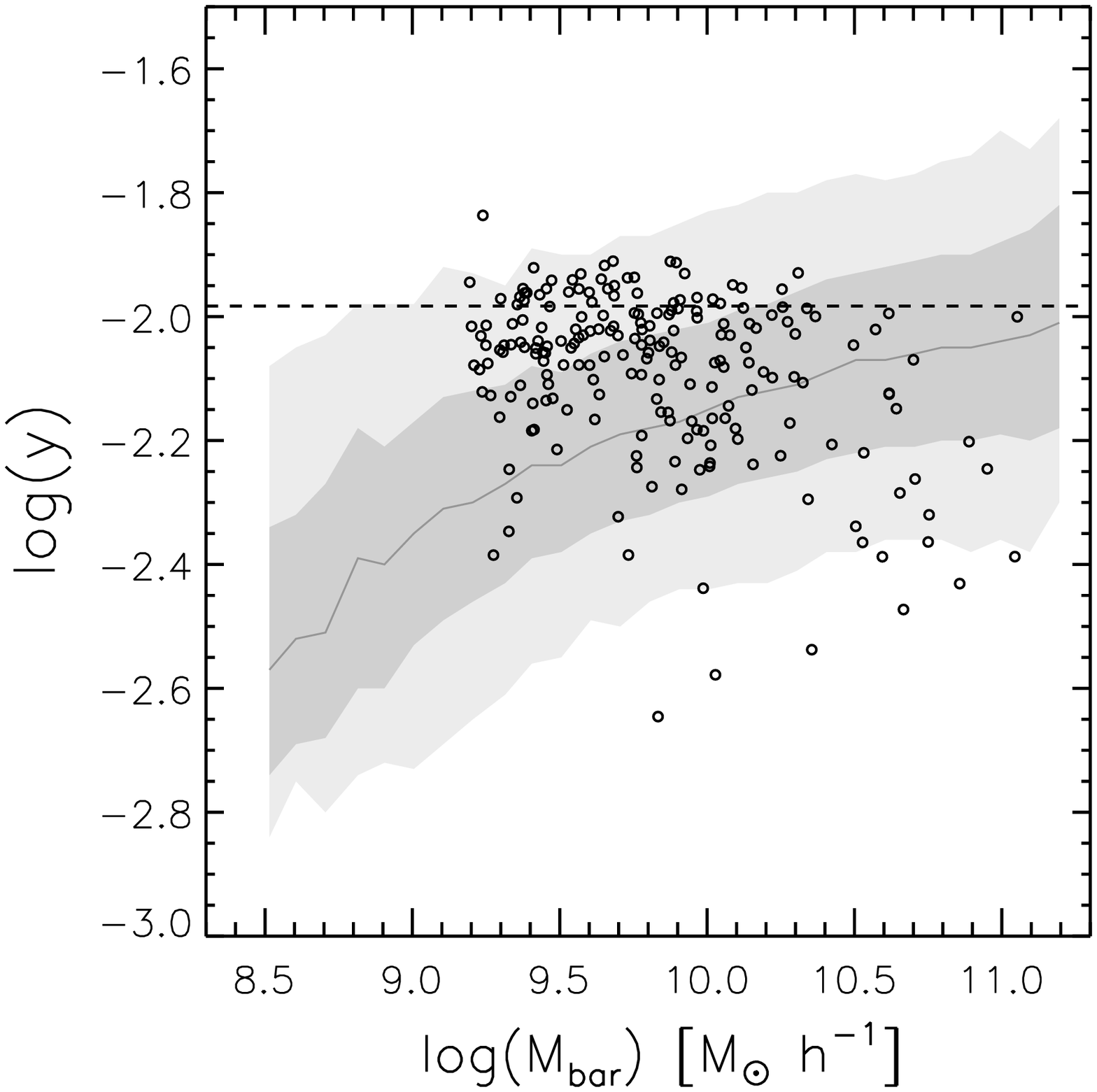}}\hspace*{-0.2cm}\\
\end{center}
\caption
{Effective yields  as a function of baryonic mass for the simulated galaxies at $z=0$. 
The shaded area corresponds to the estimations of Tremonti et al. (2004) for galaxies in SDSS.
The dashed line represents the solar value.
}
\label{yields}
\end{figure}

The need for  SN outflows is also suggested by the analysis of the effective yields
(defined as $y_{Z}= Z /{\rm ln} (\mu)^{-1}$ where $Z$ is the gas-phase metallicity and $\mu$ the gas fraction
of the systems), as shown in Fig.~\ref{yields} for the simulated galaxies at $z=0$.
These abundances were estimated within ${\rm r_{\rm apt}}$ in order
to mimic the aperture effects previously discussed. 
 The   simulated $y_{Z}$ values are lower than the solar yield
expected in a closed box model, since our systems formed in a hierarchical scenario where mergers,
 interactions and infall
affect the mass distribution and regulate star formation.
Contrary to observations (e.g. Garnett 2002), we found  larger $y_{\rm Z}$ 
for systems with $10^9 {\rm M_{\odot}} h^{-1} < M_* < M_{\rm c}$
 compared to those of the massive ones ($ M_* > M_{\rm c}$),
 which supports the claim for stronger SN outflows for the former.
In the case of massive systems, we get
a mean  of $y_{Z}$ which remains approximately constant with optical velocity.
Although this latest result is in agreement with Garnett (2002), it might be  indicating
 the need for some ejection of  material also  in massive systems in the light of
the new observational  findings of  T04, who claimed to find
a weak trend for all systems.

\section{Conclusions}

In this paper, we studied the dynamical, astrophysical and chemical properties of simulated galaxies 
in a hierarchical clustering
scenario consistent with the concordance $\Lambda$-CDM model. 
In this scenario, systems grew by the aggregation of smaller ones,
generating correlations between luminosity, stellar mass, circular velocity and metallicity
which are in general agreement with observations. 
Our main goal was to underpin the effect of the hierarchical assembly of galaxy building blocks on 
the metallicity fundamental relations.
The non-linear formation of the structure makes complex the understanding of how star formation, gas cooling and accretion,
mergers and environmental effects work together and of the impact they have on the determination of relations
between dynamical and metallicity properties.  

We found that the cooling and transformation of gas into stars inside substructures that grow hierarchically
results in the determination of a characteristic mass which segregates systems dominated, on average,  by old, red 
and metal-rich stellar populations from those more gas-rich and dominated by young stars.
The characteristic mass arises because the formation of large systems  is produced by 
the fusion of substructures dominated by old stellar populations
with a small fraction of new born stars associated to the merger events.
This behaviour generates  a change in the slope of the MZR since the outcomes of mergers are larger systems with
approximately the same mean abundances.
On the other hand, less massive systems tend to form their stars in
a more passive way or by rich-gas mergers leading to a more
strong correlation between stellar mass and metallicity. 
 This characteristic stellar mass agrees with that obtained by Kauffmann et al. (2004) from the
analysis of galaxies in the SDSS-DR2. Neverthless, the simulated evolution  of the MZR
is milder than present observational evidence. Strong SN energy feedback could help
to reconcile models and observations.

Our results suggest that this $M_{c}$ remains unchanged since $z=3$ playing  the role of segregating galaxies
into these two distinctive populations. Systems with stellar masses smaller than $M_{c}$ are the ones experiencing
the largest changes in their astrophysical properties since $z\approx3$.
These systems are responsible for the evolution of the luminosity-metallicity relation so that at a given metallicity,
simulated systems are found to be around 3 magnitudes brighter at high redshift.
Small simulated galaxies constitute a diverse population which has formed approximately 50 per cent
of their stars since $z\approx 1 $. Conversely,  large systems had half their stars formed at $z > 2$, on average.

Our results suggest that in the concordance cosmology, the hierarchical aggregation of the structure, which
regulates the gas infall and the star formation activity in the absence of energy feedback,
 is a key process which can
dynamically determine a characteristic stellar mass $M_c$. The simulated   $M_c$  plays a similar role to
that of the characteristic mass obtained from observations.
Certainly, SN feedback is also expected to play an important role in shaping the MZR as already
shown by previous works such as Larson (1974) and Chiosi \& Carraro (2002). However, it is accepted 
that the effects of SN feedback depend on the potential well of the systems. 
In hierarchical scenarios, this is an important aspect since 
the formation of galaxies may have involved the aggregation of substructures with a wide range of masses, each of
them susceptible to be differentially affected  by SN energy feedback (depending on their virial masses).
Then, to probe the effects of SN feedback, it is of upmost importance which  histories of formation are  assumed for galaxies.
In this work, we provide  galaxy formation histories consistent with the concordance models. Within this context,
  our  findings suggest that since $z=3$, SN feedback  might have a weak effect on  $M_{c}$,
 owing to the fact that this characteristic mass corresponds to
fast rotating systems even at $z=0$.
However, this point will be adequately discuss in a forthcoming paper.

\section*{Acknowledgements}

The authors are grateful to the anonymous referee for the detailed and thorouful comments that help to improve this paper and 
to A. Gallazzi for making available their data in electronic form.
 This work was supported in part by Consejo Nacional de Investigaciones Cient\'{\i}ficas y T\'ecnicas,
Agencia de Promoci\'on de Ciencia y Tecnolog\'{\i}a and Fundaci\'on Antorchas from Argentina.
 PBT thanks the Aspen Center for Physics
during the 2004 Summer Workshop for useful discussions that help to the development of this work. 
  The simulations were
performed on Ingeld PC Cluster hosted by the Numerical Astrophysics group at Institute of 
Astronomy and Space Physics. This work was partially supported by the European Union's ALFA-II
programme, through LENAC, the Latin American European Network for
Astrophysics and Cosmology.

~

\end{document}